\documentclass[reprint,aps,prb,amsmath,amssymb]{revtex4-2}
\usepackage{mathtools, bm, xcolor, graphicx, hyperref}

\usepackage{multirow}

\usepackage[capitalise]{cleveref}

\usepackage{enumitem}

\graphicspath{ {./figs/} }

\begin{document}

\title{Spin-$0$ Mott insulator to metal to spin-$1$ Mott insulator transition in the single-orbital Hubbard model on the decorated honeycomb lattice}
\author{H. L. Nourse}
\affiliation{School of Mathematics and Physics, The University of Queensland, Brisbane, Queensland 4072, Australia}
\author{Ross H. McKenzie}
\affiliation{School of Mathematics and Physics, The University of Queensland, Brisbane, Queensland 4072, Australia}
\author{B. J. Powell}
\affiliation{School of Mathematics and Physics, The University of Queensland, Brisbane, Queensland 4072, Australia}

\begin{abstract}
We study the interplay of strong electron correlations and intra-triangle spin exchange at two-thirds filling of the single-orbital Hubbard model on the decorated honeycomb lattice using rotationally invariant slave bosons. We find that the spin exchange tunes between a spin-$1$ Mott insulator, a metal, and a spin-$0$ Mott insulator when the exchange is antiferromagnetic. The Mott insulators occur from effective intra-triangle multi-orbital interactions and are adiabatically connected to the ground state of an isolated triangle. An antiferromagnetic spin exchange, as determined by the Goodenough-Kanamori rules, may occur in coordination polymers from kinetic exchange via the ligands. We characterize the magnetism in the regime where spin-triplets dominate. For small $U$ a spin-$1$ Slater insulator occurs with antiferromagnetic order between triangles. Magnetism in the spin-$1$ Mott insulator is described by a spin-$1$ Heisenberg model on a honeycomb lattice, whose ground state is N\'{e}el ordered.
\end{abstract}

\maketitle

\section{Introduction}

The Hubbard model is a paradigm for investigating strongly correlated electron systems \cite{Imada1998}. Despite the simplicity of the model - a single-orbital model on a lattice with an on-site Coulomb repulsion - it captures a plethora of correlated properties due to the competition between the kinetic energy and Coulomb repulsion. Of central importance to strongly correlated systems is the Mott metal-insulator transition \cite{Mott1968}, where at half-filling of the Hubbard model an insulator occurs with an electron localized to a lattice site. The prevailing wisdom is that away from half-filling there is instead a metal.

It has been argued that the single-orbital Hubbard model may hold all of the important ingredients for describing high-temperature superconductivity found in the hole doped cuprates \cite{Zhang1988,Zhang1990} and the organic BEDT-TTF superconductors \cite{Powell2011}. However, a single electronic band at the Fermi energy is not typical. Many strongly correlated systems are multi-orbital and originate from $d$- or $f$-shell atoms, as is found in most transition metals \cite{Tokura2000} and the iron-based superconductors \cite{Si2016}. Multi-orbital systems are typically modeled with the Hubbard-Kanamori Hamiltonian \cite{Kanamori1963,Georges2013}. In this model there are many more possibilities because of the increased orbital degrees of freedom. For example, a correlated insulator may be found away from half-filling, with a Hund's interaction that is found to enhance the critical interaction strength of the Mott insulating state \cite{Georges2013}.

\begin{figure}
	\centering
	\includegraphics[width=\columnwidth]{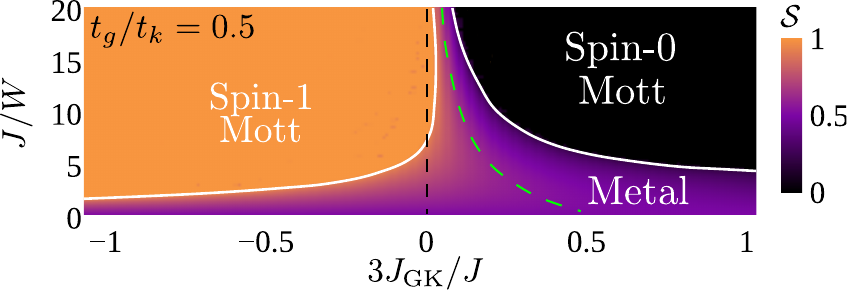}
	\caption{\label{fig:para-effective-J}
		Effective spin $\mathcal{S}$ on a triangle. An intra-triangle spin-exchange $J_{\mathrm{GK}}$ tunes between a spin-$1$ Mott insulator, a metal, and a spin-$0$ Mott insulator on the decorated honeycomb lattice at two-thirds filling.
		For $J_{\mathrm{GK}} \gtrsim 0.04$ an insulator occurs that forms a total spin-singlet on each triangle, while for $J_{\mathrm{GK}} \lesssim 0.04$ there is a net effective Hund's interaction that drives spin-triplet formation.
		The total spin $\mathcal{S}$ of a triangle is the solution to $\mathcal{S}(\mathcal{S} + 1) = \sum_i \langle \vec{S}_i \cdot \vec{S}_i \rangle / 2 \mathcal{N}$, where the spin operator of triangle $i$ is $\vec{S}_i = \sum_{\alpha = 1}^3 \sum_{\sigma \sigma'} \hat{c}_{i\alpha\sigma}^{\dagger} \vec{\bm{\tau}}_{\sigma \sigma'} \hat{c}_{i\alpha\sigma'}$, $\vec{\bm{\tau}}$ is a vector of Pauli matrices, and $\mathcal{N}$ is the number of unit cells on the lattice.
		$J = U + 2J_{\mathrm{GK}}$, the green dashed line marks the boundary between spin-$0$ and spin-$1$ ground states of an isolated triangle (cf.~\cref{fig:isolated-effective-J,fig:para-effective-J-energy}), the black dashed line marks $J_{\mathrm{GK}} = 0$, and the solid white lines mark the phase boundaries.
	}
\end{figure}

However, increasing the orbital degrees of freedom is not the only way to increase the complexity of strongly correlated systems. For example, some coordination polymers have elaborate lattices \cite{Murase2017,Murase2017b,Kingsbury2017,Jeon2015,Darago2015,DeGayner2017,Henling2014,Henline2014,Polunin2015,Kalmutzki2018}. Many coordination polymers display properties observed in strongly correlated systems, such as Kondo physics \cite{Jiang2019,Kumar2021}, and unconventional superconductivity \cite{Zhang2017,Huang2018,Takenaka2021}. Recently, it was shown that decorated lattices support a plethora of strongly correlated phases away from half-filling, despite being described by the single-orbital Hubbard model \cite{Nourse2021}. One of these surprising states was a Mott insulator on the decorated honeycomb lattice with spin-triplet formation, denoted as a spin-$1$ Mott insulator.

Only the screened on-site Coulomb repulsion $U$ is kept in the Hubbard model. However, in many materials, such as coordination polymers, there may be important spin exchange processes, which we will show to have profound effects. As the spin-$1$ Mott insulator occurs in the regime of strong intra-triangle hopping on the decorated honeycomb lattice, we investigate the effect of an intra-triangle spin exchange $J_{\mathrm{GK}}$. The sign of the spin exchange determines whether ferromagnetic (negative) or antiferromagnetic (positive) spin configurations are favored between adjacent sites. In \cref{sec:gka-rules} we discuss how $J_{\mathrm{GK}}$ is estimated in materials with metals bridged by ligands, such as coordination polymers, where the Goodenough-Kanamori-Anderson (GKA) rules are often used.

Our central result is summarized in \cref{fig:para-effective-J}. An antiferromagnetic spin exchange $J_{\mathrm{GK}}$ suppresses spin triplet formation on a triangle and there is instead a spin-$0$ Mott insulator. It is possible to tune between a spin-$1$ Mott insulator, a metal, and a spin-$0$ Mott insulator by tuning $J_{\mathrm{GK}}$. Furthermore, a ferromagnetic $J_{\mathrm{GK}}$ highlights that the spin-$1$ Mott insulator occurs from an effective Hund's rule coupling and is connected to physics found in the Hubbard-Kanamori model.

The paper is organized as follows: The extended Hubbard model on the decorated honeycomb lattice is introduced in \cref{sec:model} and presented in the basis of molecular orbitals of a triangle, which we denote the trimer orbitals. We discuss three limits where our model simplifies: the non-interacting limit (no onsite Coulomb repulsion $U$ nor $J_{\mathrm{GK}}$), finite $U$ and $J_{\mathrm{GK}}$ but isolated triangles (inter-triangle hopping $t_g = 0$, intra-triangle hopping $t_k > 0$), and finite $U$ and $J_{\mathrm{GK}}$ but infinitely separated in energy trimer orbitals ($t_g \neq 0$, $t_k \rightarrow \infty$). Mean-field rotationally invariant slave bosons (RISB) are introduced in \cref{sec:risb}, which is the method we use to solve the model. In \cref{sec:para} we present our results restricted to paramagnetic states and characterize in detail the metal-insulator transition to a Mott insulator with spin-$1$ formation. In \cref{sec:spin-state-transitions} we discuss how an intra-triangle spin exchange can suppress triplet formation on a triangle and instead drive a spin-$0$ Mott insulator. In \cref{sec:afm} we discuss antiferromagnetic solutions, where a spin-density wave forms between spin-triplet polarized triangles. In \cref{sec:low-energy-theory} we discuss what the magnetism looks like in the spin-$1$ Mott insulator. Finally, \cref{sec:conclusion} presents our conclusions and the implication of our results for other decorated lattices.

\section{The model} \label{sec:model}

\begin{figure}
	\centering
	\includegraphics[width=\columnwidth]{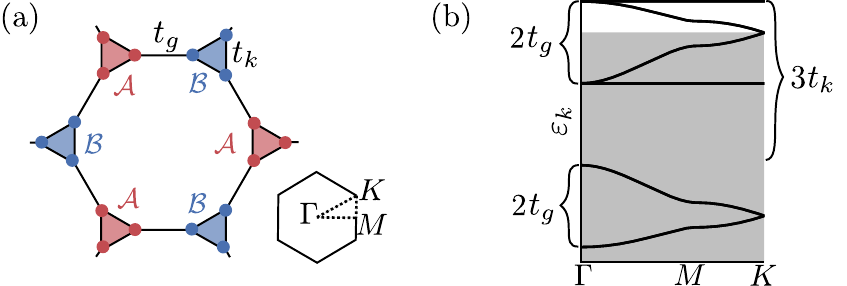}
	\caption{\label{fig:lattice}
		(a) The decorated honeycomb lattice. The intra-(inter)triangle hopping amplitude is $t_k$ ($t_g$), and $\mathcal{A}$ and $\mathcal{B}$ label the inequivalent triangles in the unit cell. The inset shows the hexagonal Brillouin zone. $\Gamma$, $K$, and $M$ are the $k$-points of high symmetry in reciprocal space.
		(b) Non-interacting band structure of the Dirac metal at two-thirds filling for $t_g/t_k = 0.5$. The dispersion $\varepsilon_k$ is shown between the points of high symmetry, and the shaded region indicates electron filling up to the Fermi energy.
	}
\end{figure}

The Hamiltonian for the Hubbard model with an intra-triangle spin-exchange on the decorated honeycomb lattice \cite{Jacko2015} is
\begin{align} \label{eq:ham}
\hat{H} & \equiv  \hat{H}^{\triangle \rightarrow \triangle} + \sum_i \hat{H}_i^{\triangle}, \\
\label{eq:ham-kin}
\hat{H}^{\triangle \rightarrow \triangle} & \equiv - t_g \sum_{\langle i \alpha, j \alpha \rangle, \sigma} \hat{c}_{i\alpha\sigma}^{\dagger} \hat{c}_{j\alpha\sigma}^{}, \\
\label{eq:ham-loc}
\hat{H}_i^{\triangle} & \equiv \hat{H}_i^0 + \hat{H}_i^U + \hat{H}_i^{\mathrm{GK}}, \\
\label{eq-ham-0}
\hat{H}_i^0 & \equiv - t_k \sum_{\alpha\neq\beta ,\sigma} \hat{c}_{i\alpha\sigma}^{\dagger} \hat{c}_{i\beta\sigma}^{}, \\
\label{eq:ham-U}
\hat{H}_i^U & \equiv U\sum_{\alpha} \hat{n}_{i\alpha\uparrow} \hat{n}_{i\alpha\downarrow}, \\
\label{eq:ham-F}
\hat{H}_i^{\mathrm{GK}} & \equiv J_{\mathrm{GK}} \sum_{\alpha \neq \beta} \left(\vec{S}_{i\alpha} \cdot \vec{S}_{i\beta}- \frac{\hat{n}_{i\alpha} \hat{n}_{i\beta}}{4} \right),
\end{align}
where $\hat{c}_{i\alpha\sigma}^{\dagger}$ ($\hat{c}_{i\alpha\sigma}^{}$) creates (annihilates) an electron with spin $\sigma \in \{\uparrow,\downarrow\}$ on site $\alpha \in \{1,2,3\}$ of triangle $i$, $\hat{n}_{i\alpha} \equiv \sum_{\sigma} \hat{n}_{i\alpha\sigma}$, $\hat{n}_{i\alpha\sigma} \equiv \hat{c}_{i\alpha\sigma}^{\dagger} \hat{c}_{i\alpha\sigma}^{}$, $t_g$ ($t_k$) is the inter-(intra-)triangle hopping integral (\cref{fig:lattice}a), $\langle i\alpha, j\alpha\rangle$ signifies nearest-neighbor hopping between sites of adjacent triangles, $U$ is the local Coulomb repulsion, and $J_{\mathrm{GK}}$ is the intra-triangle spin exchange.

\subsection{The molecular orbital basis for two-thirds filling}

\begin{figure}
	\centering
	\includegraphics[width=\columnwidth]{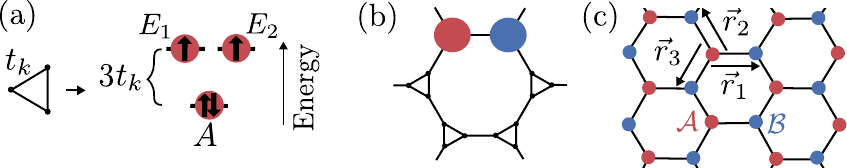}
	\caption{\label{fig:trimer-ground-state}
		(a) The trimer orbitals of an equilateral triangle. For $t_k >0$ the $A$ molecular orbital has lower energy than the two degenerate $E$ molecular orbitals. The ground state of the single-orbital Hubbard model ($J_{\mathrm{GK}} = 0$) of a three-site ring at two-thirds electron filling and $U>0$ is a spin-triplet.
		(b) In the basis of the trimer orbitals the decorated honeycomb lattice maps to a three-orbital model on the
		(c) honeycomb lattice with complicated phase dependent inter- and intra-orbital hopping between trimers. The vectors to nearest neighbor sites is given by $\vec{r}_1 = (\vec{a}_1 + \vec{a}_2) / 3$, $\vec{r}_2 = (\vec{a}_1 - 2 \vec{a}_2) / 3$, and $\vec{r}_3 = (\vec{a}_2 - 2 \vec{a}_1)/3$, where the primitive lattice vectors are $\vec{a}_1 = a(3/2, \sqrt{3}/2)$ and $\vec{a}_2 = a(3/2, -\sqrt{3}/2)$, where $a = 1$ is the lattice spacing.
	}
\end{figure}

It will be helpful below to consider the molecular orbitals of a triangle, which we denote as the trimer orbitals. To do so we write
\begin{equation}
\hat{c}_{i\alpha\sigma}^{\dagger} \equiv \frac{1}{\sqrt{3}} \sum_m \hat{b}_{im\sigma}^{\dagger} e^{-i \phi(\alpha - 1)m},
\end{equation}
where $\phi = 2\pi/3$, $m \in \{ -1,0,1 \} \equiv \{E_1, A, E_2\}$ labels the trimer orbitals. The intra-triangle terms (\cref{eq:ham-loc}) have the representation (\cref{sec:app-local,sec:app-hunds})
\begin{align} 
\label{eq-ham-0-trimer}
& \hat{H}_i^{0} = -2t_k \sum_{m,\sigma} \cos(\phi m) \hat{b}_{im\sigma}^{\dagger} \hat{b}_{im\sigma}^{}, \\
\label{eq-ham-U-trimer}
& \hat{H}_i^U = - \frac{U}{3} \vec{S}_i^2 + \frac{U}{12} \hat{N}_i^2 + \frac{U}{6} \hat{N}_i - \frac{U}{3} \sum_m \hat{n}_{im\uparrow} \hat{n}_{im\downarrow} \nonumber \\
& + \frac{U}{3} \sum_{m} \sum_{n \neq m} \sum_{p \neq m \neq n} \left( \hat{b}_{im \downarrow}^{\dagger} \hat{b}_{im \uparrow}^{\dagger} \hat{b}_{in \uparrow}^{} \hat{b}_{ip \downarrow}^{} + \mathrm{H.c.} \right), \\
\label{eq-ham-F-trimer}
& \hat{H}_i^{\mathrm{GK}} = \frac{J_{\mathrm{GK}}}{3} \hat{S}_i^2 - \frac{J_{\mathrm{GK}}}{12} \hat{N}_i^2 - \frac{J_{\mathrm{GK}}}{6} \hat{N}_i - \frac{2 J_F}{3} \sum_{m} \hat{n}_{im\uparrow} \hat{n}_{im\downarrow} \nonumber \\
& + \frac{2 J_{\mathrm{GK}}}{3} \sum_{m} \sum_{n \neq m} \sum_{p \neq m \neq n} \left( \hat{b}_{im \downarrow}^{\dagger} \hat{b}_{im \uparrow}^{\dagger} \hat{b}_{in \uparrow}^{} \hat{b}_{ip \downarrow}^{} + \mathrm{H.c.} \right),
\end{align}
where $\hat{n}_{im \sigma} \equiv \hat{b}_{im\sigma}^{\dagger} \hat{b}_{im\sigma}^{}$, $\hat{N}_i \equiv \sum_{m,\sigma} \hat{n}_{i m \sigma}$, $\vec{S}_i^2 = \vec{S}_i \cdot \vec{S}_i$, $\vec{S}_i = \sum_m \vec{S}_{im}$, and $\vec{S}_{im} = (S_{im}^x, S_{im}^y, S_{im}^z)$ is given by
\begin{equation}
\vec{S}_{im} \equiv \frac{1}{2} \sum_{\sigma \sigma'} \hat{b}_{im \sigma}^{\dagger} \vec{\bm{\tau}}_{\sigma \sigma'} \hat{b}_{im \sigma'}^{},
\end{equation}hat
where $\vec{\bm{\tau}}$ is a vector of Pauli matrices. $\hat{H}_i^0$ (\cref{eq-ham-0-trimer}) defines the energy of the trimer orbitals, with the degenerate $E_1$ and $E_2$ orbitals $3t_k$ higher in energy than the $A$ orbital for $t_k >0$ (\cref{fig:trimer-ground-state}a). One can can infer the effect of the interactions by considering the case $J_{\mathrm{GK}} = 0$. The $\hat{N}_i^2$ term disfavors charge fluctuations on a triangle. The $\vec{S}_i^2$ term favors maximizing the total spin on a triangle and reflects Hund's first rule. The intra-orbital Coulomb attraction favors low spin configurations and competes with the $\vec{S}_i^2$ term. The last term in $\hat{H}_i^U$ and $\hat{H}_i^{\mathrm{GK}}$ (\cref{eq-ham-U-trimer,eq-ham-F-trimer} respectively) describes a three-orbital hopping process that creates (annihilates) double occupancy on an orbital. See \cref{sec:app-local,sec:app-hunds} for details.

At two-thirds filling the Hubbard model of an isolated triangle is exactly solvable and is exact within RISB with triangular clusters. For $t_k >0$, $U > 0$, $J_{\mathrm{GK}} = 0$ the ground state is three-fold degenerate with energy 
\begin{equation} \label{eq:energy-isolated-triangle}
E_0^{\triangle}(J_{\mathrm{GK}} = 0) = -2t_k + U
\end{equation}
and 
%
%
the three triplet states are \cite{Janani2014a}
\begin{align}
	|t_i^{\Uparrow} \rangle & \equiv \hat{b}_{iA\uparrow}^{\dagger} \hat{b}_{iA\downarrow}^{\dagger} \hat{b}_{iE_1\uparrow}^{\dagger} \hat{b}_{iE_2\uparrow}^{\dagger} |0 \rangle, \nonumber \\
	|t_i^0 \rangle& \equiv \frac{1}{\sqrt{2}} \hat{b}_{iA\uparrow}^{\dagger} \hat{b}_{iA\downarrow}^{\dagger} \left( \hat{b}_{iE_1\uparrow}^{\dagger} \hat{b}_{iE_2\downarrow}^{\dagger} + \hat{b}_{iE_1\downarrow}^{\dagger} \hat{b}_{iE_2\uparrow}^{\dagger} \right) |0 \rangle, \nonumber \\
	|t_i^{\Downarrow} \rangle & \equiv \hat{b}_{iA\uparrow}^{\dagger} \hat{b}_{iA\downarrow}^{\dagger} \hat{b}_{iE_1\downarrow}^{\dagger} \hat{b}_{iE_2\downarrow}^{\dagger} |0 \rangle,
\end{align}
where $|0\rangle$ is the vacuum. A triplet ground state occurs on an isolated triangle because it allows the one minority spin to minimize its kinetic energy \cite{Janani2014a}.

The inter-triangle hopping $t_g$ couples the molecular orbitals of adjacent triangles. In the molecular orbital basis \cref{eq:ham-kin} can be written as
\begin{equation}
\hat{H}^{\triangle \rightarrow \triangle} = \sum_m \hat{H}_m^{\mathrm{honeycomb}} + \sum_{m\neq n} \hat{H}_{mn}^{\mathrm{trimer}},
\end{equation}
which is a complicated three-orbital tight-binding model on a honeycomb-like lattice (\cref{fig:trimer-ground-state}b,c). The first term is the tight-binding model on the honeycomb lattice, given in reciprocal space by 
\begin{equation}
	\hat{H}_m^{\mathrm{honeycomb}} = \frac{1}{3} \sum_{k\sigma} \vec{\Psi}_{km\sigma}^{\dagger} \bm{\mathcal{H}}_{k}^{\mathrm{honeycomb}} \vec{\Psi}_{km\sigma}^{},
\end{equation}
with $\vec{\Psi}_{km\sigma} \equiv (\hat{b}_{\mathcal{A}km\sigma}, \hat{b}_{\mathcal{B}km\sigma})^{\mathrm{T}}$,
\begin{align}
	\bm{\mathcal{H}}_k^{\mathrm{honeycomb}} = - t_g \begin{pmatrix}
		0 & \Delta_k \\
		\Delta_k^* & 0
	\end{pmatrix},
\end{align}
where $k$ indexes a vector $\vec{k}$ in reciprocal space, $\ell = \mathcal{A}, \mathcal{B}$ in $\hat{b}_{\ell km\sigma}$ labels the two sites in the unit cell of the honeycomb lattice (\cref{fig:lattice}a), and $\Delta_k = \sum_{j=1}^{3} e^{i \vec{k} \cdot \vec{r}_j}$ with $\vec{r}_j$ denoting the three vectors to the nearest-neighbor sites in real-space (\cref{fig:trimer-ground-state}c).
The second term between different trimer orbitals takes a similar form but has a direction dependent phase, given by
\begin{equation}
	\hat{H}_{mn}^{\mathrm{trimer}} = \frac{1}{3} \sum_{k\sigma} \vec{\Psi}_{km\sigma}^{\dagger} \bm{\mathcal{H}}_{k}^{\mathrm{trimer}} \vec{\Psi}_{kn\sigma}^{},
\end{equation}
with
\begin{align}
	\bm{\mathcal{H}}_k^{\mathrm{trimer}} = - t_g \begin{pmatrix}
		0 & \delta_k \\
		\delta_k^* & 0
	\end{pmatrix},
\end{align}
where $\delta_k = \sum_{j=1}^{3} e^{i (\vec{k} \cdot \vec{r}_j + (j-1) \phi)}$.

\subsection{The non-interacting limit and the limit of infinitely separated bands} \label{sec:non-interacting-limit}

In \cref{fig:lattice}b we show an example of the non-interacting ($U = 0$, $J_{\mathrm{GK}} = 0$) band structure for $t_g/t_k < 3/2$. At $\vec{k}=\vec{0}$ (the $\Gamma$ point) the eigenstates of \cref{eq:ham} can be labeled by the trimer orbitals and the (anti-)bonding orbitals of the honeycomb lattice. Hence, at $\Gamma$, the (anti-)bonding orbitals cause the two $A$ eigenenergies to be separated in energy by $2t_g$, and the four $E$ eigenenergies that are two-fold degenerate to be separated in energy by $2t_g$. The set of $A$ eigenenergies are $3t_g$ lower in energy than the set of $E$ eigenenergies because of the energy difference between the molecular orbitals of an isolated triangle. Away from $\Gamma$ the labeling is not exact, but the set of lower (upper) bands retain significant $A$ ($E$) orbital character and are honeycomb-like (\cref{fig:lattice}b).

In the limit $t_g \neq 0$, $t_k \rightarrow \infty$, and $U = J_{\mathrm{GK}} = 0$ the $A$ and $E$ orbitals are infinitely separated in energy and the upper set of bands decouple from the lower set. For finite $U$, the three-orbital hopping terms in \cref{eq-ham-U-trimer,eq-ham-F-trimer} can be neglected because it always involves moving an electron (hole) from the $A$ ($E$) bands to the $E$ ($A$) bands, which costs infinite energy. Electrons in the $A$ orbitals are exactly described by the Hubbard model with the intra-triangle spin-exchange on the honeycomb lattice. Electrons in the $E$ orbitals are exactly described by a degenerate two-orbital Hubbard-Kanamori model on a honeycomb-like lattice with phase dependent hopping. 

\section{Methods: RISB mean-field theory} \label{sec:risb}

We solve \cref{eq:ham} using mean-field rotationally invariant slave-boson (RISB) theory \cite{Kotliar1986,Lechermann2007,Lanata2015,Lanata2017} using three-site (triangular) clusters. At the saddle-point level RISB captures the coherent low-energy quasiparticles in metallic states, and captures metal-insulator transitions in the paramagnetic state of multi-orbital systems. Importantly, rotational invariance allows us to work in the molecular orbital basis of a triangle.  

In RISB the local physical Hilbert space of a cluster is mapped onto a larger Hilbert space described by auxiliary fermions and bosons. The advantage is that the auxiliary fermions only enter the problem quadratically and can be integrated out, at the expense of an action that depends on the slave bosons and Lagrange multiplier fields, which are evaluated at the saddle-point level. 

A new set of auxiliary fermions are introduced for each electronic mode (sites, orbitals, and spin) on a cluster $\{\hat{f}_{ia} | a = 1, ..., M_i \}$, where $M_i = 6$ is the number of electronic modes on a triangle. There are $2^{M_i}$ physical states $\{|A_i\rangle \}$ and $2^{M_i}$ auxiliary fermion states $\{|n_i \rangle \}$ on a cluster. A boson $\hat{\Phi}_{iAn}$ is introduced for each pair of physical and auxiliary fermion states $\{ |A_i \rangle, |n_i \rangle \}$, with the restriction that the number of particles in state $|A_i \rangle$, denoted as $N_{iA}$, is equal to the number of particles in state $|n_i \rangle$, denoted as $N_{in}$.

In the enlarged Hilbert space, a faithful representation of the physical electron is given by ($i$ indexes a cluster and $\alpha$ and $a$ index sites, orbitals, and spin)
\begin{equation} \label{eq:sb-electron}
	\underline{\hat{c}}_{i\alpha}^{\dagger} \equiv \sum_a \hat{\mathcal{R}}_{ia\alpha}^{} \hat{f}_{ia}^{\dagger},
\end{equation}
with the unitary operator
\begin{equation}
\hat{\mathcal{R}}_{ia\alpha} \equiv \sum_{AB} \sum_{nm} \frac{\langle A| \hat{c}_{i\alpha}^{\dagger} |B\rangle \langle n| \hat{f}_{ia}^{\dagger} | m \rangle }{\sqrt{N_{iA}(M_i - N_{iB})}} \hat{\Phi}_{iAn}^{\dagger} \hat{\Phi}_{iBm}^{}.
\end{equation}

To restrict the enlarged Hilbert space to only the physical states the following (Gutzwiller) constraints are introduced
\begin{align}
\hat{K}_i^0 & \equiv \sum_{An} \hat{\Phi}_{iAn}^{\dagger} \hat{\Phi}_{iAn}^{} - \hat{1}, \\
\hat{K}_{iab} & \equiv \hat{f}_{ia}^{\dagger} \hat{f}_{ib}^{} - \sum_{Anm} \langle m_i | \hat{f}_{ia}^{\dagger} \hat{f}_{ib}^{} | n_i \rangle \hat{\Phi}_{iAn}^{\dagger} \hat{\Phi}_{iAm},
\end{align}
where $\hat{1}$ is the identity. $\hat{K}_i^0$ enforces that only states with a single boson per cluster are retained, while $\hat{K}_{iab}$ enforces that a physical state has the correct bosons attached to their corresponding auxiliary fermion. The constraints are incorporated into the Hamiltonian (\cref{eq:ham}) in the enlarged Hilbert space with Lagrange multipliers $E_i^c$ and $\lambda_{iab}$ respectively.

With the above restrictions, a physical state in the enlarged Hilbert space is given by
\begin{equation}
|\underline{A}_{i} \rangle \equiv \frac{1}{\sqrt{D_{iA}}} \sum_n \hat{\Phi}_{iAn}^{\dagger} |0\rangle \otimes |n_i \rangle,
\end{equation}
where $D_{iA} = {M_i \choose N_{iA}}$ is the number of auxiliary fermion states with $N_{iA}$ particles. Hence, one can show that any local observable acting on a cluster in the enlarged Hilbert space has a faithful representation given by \cite{Lechermann2007}
\begin{equation}
	\hat{X} [ \underline{c}_{i\alpha}^{\dagger}, \underline{c}_{i\alpha}^{} ] = \sum_{AB} \langle A | \hat{X} [ \hat{c}_{i\alpha}^{\dagger}, \hat{c}_{i\alpha}^{} ] | B \rangle \sum_n \hat{\Phi}_{iAn}^{\dagger} \hat{\Phi}_{iBn}^{}.
\end{equation}

At the saddle-point level the bosons condense ($\langle \hat{\Phi}_{iAn} \rangle \rightarrow [\phi_i]_{An}$, $\langle \hat{\mathcal{R}}_{ia\alpha} \rangle \rightarrow [\bm{\mathcal{R}}_i]_{a\alpha}$), and the constraints are only enforced on average. The local density matrix $\bm{\Delta}_i^p$ of $\hat{H}^{\mathrm{qp}}$ (\cref{eq:ham-qp}) and the renormalization matrix $\bm{\mathcal{R}}_i$ are promoted to free parameters with the Lagrange-Legendre terms
\begin{align}
[\bm{\lambda}_i^c]_{ab} \left( \sum_{Anm} \langle n| \hat{f}_{ia}^{\dagger} \hat{f}_{ib} |m \rangle [\phi_i^{\dagger}]_{nA} [\phi_i^{}]_{Am} - [\Delta_i^p]_{ab} \right), \\
[\bm{\mathcal{D}}_i]_{a\alpha} \left( \sum_{ABnm} \langle A |\hat{c}_{i\alpha}^{\dagger} | B \rangle \langle m | \hat{f}_{ia} |n\rangle [\phi_i^{\dagger}]_{nA} [\phi_i^{}]_{Bm} \right. \nonumber \\
\left. -\sum_c [\mathcal{R}_i]_{c\alpha} [ \Delta_i^p(1-\Delta_i^p)]_{ca}^{1/2} \right) + \mathrm{H.c.}
\end{align}
where their definitions are enforced with the Lagrange multiplier fields $[\bm{\lambda}_i^c]_{ab}$ and $[\bm{\mathcal{D}}_i]_{a\alpha}$ respectively \cite{Lanata2015,Lanata2017}. The resulting mean-field theory is entirely encoded by the Lagrange functional \cite{Lanata2017b}
\begin{align} \label{eq:risb-grand-potential}
\mathcal{L}[\bm{\mathcal{R}}_i, & \bm{\lambda}_i, \bm{\Delta}_i^p, E_i^c, \bm{\mathcal{D}}_i, \bm{\lambda}_i^c,|\Phi_i\rangle] \nonumber \\
&=  \lim_{\beta\rightarrow \infty} \frac{1}{\beta} \sum_{i\omega_n} \mathrm{Tr} \ln \hat{G}^{\mathrm{qp}}(i\omega_n) \nonumber \\
& + \sum_i \left[ \langle \Phi_i | \hat{H}_i^{\mathrm{emb}} | \Phi_i \rangle + E_i^c \left( 1 - \langle \Phi_i | \Phi_i \rangle \right) \right] \nonumber \\
& - \sum_{i,a\alpha c} \left( [\bm{\mathcal{D}}_i]_{a\alpha} [\bm{\mathcal{R}}_i]_{c\alpha} [\bm{\Delta}_i^p(1-\bm{\Delta}_i^p)]_{ca}^{1/2} + \textrm{c.c.} \right) \nonumber \\
& -\sum_{i, ab} \left( [\bm{\lambda}_i^{}]_{ab} + [\bm{\lambda}_i^c]_{ab} \right) [\bm{\Delta}_i^p]_{ab},
\end{align}
where the auxiliary fermions Green's function is given by $\hat{G}^{\mathrm{qp}}(z) \equiv (z - \hat{H}^{\mathrm{qp}})^{-1}$ with the Hamiltonian
\begin{align} \label{eq:ham-qp}
\hat{H}^{\mathrm{qp}} &\equiv -\sum_{ij,\alpha\beta,ab} [\bm{\mathcal{R}}_i^{}]_{a\alpha}^{} [t_{ij}]_{\alpha\beta}  [\bm{\mathcal{R}}_j^{\dagger}]_{\beta b}^{} \hat{f}_{ia}^{\dagger} \hat{f}_{jb}^{} \nonumber \\
& \phantom{{}\equiv} + \sum_{i,ab} [\bm{\lambda}_i^{}]_{ab}^{} \hat{f}_{ia}^{\dagger} \hat{f}_{ib}^{}, 
\end{align}
$|\Phi_i\rangle$ is the ground state of the embedding Hamiltonian which is an impurity problem given by the Hamiltonian
\begin{align} \label{eq:ham-emb}
\hat{H}_i^{\mathrm{emb}} & \equiv \hat{H}_i^{\triangle} + \sum_{a\alpha} \left( [\bm{\mathcal{D}}_i]_{a\alpha} \hat{c}_{i\alpha}^{\dagger} \hat{f}_{ia}^{} + \mathrm{H.c.} \right) \nonumber \\
& + \sum_{ab} [\bm{\lambda}_i^c]_{ab} \hat{f}_{ib}^{} \hat{f}_{ia}^{\dagger},
\end{align}
where $\bm{\lambda}_i^c$ and $\bm{\mathcal{D}}_i$ are matrices that describe the bath and the hybridization between the bath and the impurity respectively, and $E_i^c$ enforces the normalization of $|\Phi_i \rangle$. The same grand potential can be derived from the Gutzwiller approximation \cite{Kotliar1986,Metzner1989,Bunemann2007,Lanata2008}.

The parameters of the Lagrange functional (\cref{eq:risb-grand-potential}) have to be determined self-consistently. We assume a uniform lattice where each triangle is equivalent and drop the label $i$. The Lagrange functional is extremized with respect to $\bm{\mathcal{R}}$, $\bm{\lambda}$, $\bm{\Delta}^p$, $E^c$, $\bm{\mathcal{D}}$, $\bm{\lambda}^c$, and $|\Phi\rangle$, leading to the following equations:
\begin{align} \label{eq:selfcon-1}
[\bm{\Delta}^p]_{ab} = \frac{1}{\mathcal{N}} \left[ \sum_k f(\bm{\varepsilon}_k^{\mathrm{qp}})  \right]_{ba},
\end{align}
\begin{multline} \label{eq:selfcon-2}
[\bm{\mathcal{D}}]_{a\alpha} = \sum_c \left[ \bm{\Delta}^p(1-\bm{\Delta}^p) \right]_{ac}^{-1/2} \\ \times \left[ \frac{1}{\mathcal{N}} \sum_k \bm{\varepsilon}_k \bm{\mathcal{R}}^{\dagger} f(\bm{\varepsilon}_k^{\mathrm{qp}}) \right]_{c\alpha},
\end{multline}
\begin{multline} \label{eq:selfcon-3}
[\bm{\lambda}^c]_{ab} = -[\bm{\lambda}]_{ab} -\frac{1}{2} \left( \left[ (\bm{\mathcal{R}} \bm{\mathcal{D}})^{\mathrm{T}} \phantom{(\bm{\Delta}^p(1-\bm{\Delta}^p))^{-1/2} (1-2\bm{\Delta}^p)} \right. \right. \\ 
\left. \left.  \times (\bm{\Delta}^p(1-\bm{\Delta}^p))^{-1/2} (1-2\bm{\Delta}^p) \right]_{ba} + \mathrm{c.c.} \right),
\end{multline}
\begin{align} \label{eq:selfcon-4}
	\hat{H}^{\mathrm{emb}} |\Phi\rangle = E^c |\Phi \rangle,
\end{align}
\begin{multline} \label{eq:selfcon-5}
[\bm{\lambda}]_{ab} = -[\bm{\lambda}^c]_{ab} - \frac{1}{2} \left( \left[ ( \bm{\mathcal{R}}  \bm{\mathcal{D}})^{\mathrm{T}} (\bm{N}^f(1-\bm{N}^f))^{-1/2} \phantom{(1-2\bm{N}^f)} \right. \right. \\
\left. \left. \times  (1-2\bm{N}^f) \right]_{ba} + \mathrm{c.c.} \right),
\end{multline}
\begin{align} \label{eq:selfcon-6}
[\bm{\mathcal{R}}]_{a\alpha} = \sum_{c} [\bm{M}^{cf}]_{\alpha c} \left[ (\bm{N}^f(1-\bm{N}^f))^{-1/2} \right]_{ca},
\end{align}
where $\bm{\varepsilon}_k$ is the dispersion matrix of $\hat{H}^{\triangle\rightarrow \triangle}$ (\cref{eq:ham-kin}), $\bm{\varepsilon}_k^{\mathrm{qp}}$ is the dispersion matrix of $\hat{H}^{\mathrm{qp}}$ (\cref{eq:ham-qp}), $f(\varepsilon)$ is the Fermi function, $[\bm{N}^f]_{ab} \equiv \langle \Phi| \hat{f}_b^{} \hat{f}_a^{\dagger} | \Phi \rangle$, and $[\bm{M}^{cf}]_{\alpha a} \equiv \langle \Phi| \hat{c}_{\alpha}^{\dagger} \hat{f}_a^{} | \Phi \rangle$. The RISB saddle-point equations are solved by the following procedure: (i) guess $\bm{\lambda}$, $\bm{\mathcal{R}}$ and construct $\hat{H}^{\mathrm{qp}}$ (\cref{eq:ham-qp}); (ii) solve \cref{eq:selfcon-1,eq:selfcon-2,eq:selfcon-3} for $\bm{\Delta}^p$, $\bm{\mathcal{D}}$, and $\bm{\lambda}^c$; (iii) construct $\hat{H}^{\mathrm{emb}}$ (\cref{eq:ham-emb}) and solve the resulting Hamiltonian (\cref{eq:selfcon-4}) for the lowest energy state $|\Phi\rangle$ with $M$ fermions; (iv) evaluate $\bm{N}^f$, $\bm{M}^{cf}$ and solve \cref{eq:selfcon-5,eq:selfcon-6} for a new guess for $\bm{\lambda}$ and $\bm{\mathcal{R}}$; (v) repeat until $\bm{\lambda}$ and $\bm{\mathcal{R}}$ converge.

At the saddle-point level the physical electrons Green's function is a matrix given by
\begin{align} \label{eq:greens-fnc}
\bm{G}(k,\omega) & \equiv \bm{\mathcal{R}}^{\dagger} \bm{G}^{\mathrm{qp}}(k,\omega) \bm{\mathcal{R}}^{} \nonumber \\
& = \left[ \omega (\bm{\mathcal{R}}^{\dagger} \bm{\mathcal{R}}^{})^{-1} - \bm{\mathcal{R}}^{-1} \bm{\lambda} \bm{\mathcal{R}}^{\dagger -1} - \bm{\varepsilon}_k  \right]^{-1} \nonumber \\
& = \left[ \omega \bm{1} - \bm{\varepsilon}_k - \bm{\varepsilon}^0 - \bm{\Sigma}(\omega) \right]^{-1},
\end{align}
where the self-energy is
\begin{align} \label{eq:self-energy}
\bm{\Sigma}(\omega) = \omega(\bm{1} - (\bm{\mathcal{R}}^{\dagger} \bm{\mathcal{R}}^{})^{-1}) + \bm{\mathcal{R}}^{-1} \bm{\lambda} \bm{\mathcal{R}}^{\dagger -1} - \bm{\varepsilon}^0,
\end{align}
and $\bm{\varepsilon}^0$ are the one-body terms in $\hat{H}^{\triangle}$, \cref{eq:ham-loc}. The spectral function matrix $\bm{A}(k,\omega) \equiv -\pi^{-1} \mathrm{Im} \bm{G}(k,\omega)$ is given by
\begin{equation} \label{eq:spectral-function}
\bm{A}(k,\omega) = \bm{\mathcal{R}}^{\dagger} \delta(\omega \bm{1} - \bm{\varepsilon}_k^{\mathrm{qp}}) \bm{\mathcal{R}},
\end{equation}
so that the spectral weight in the orbitals is obtained as
\begin{equation} \label{eq:spectral-weight}
\int_{-\infty}^{\infty} \mathrm{d} \omega [\bm{A}]_{\alpha\beta}(k,\omega) = [\bm{R}^{\dagger} \bm{R}^{}]_{\alpha \beta} =  [\bm{Z}]_{\alpha\beta},
\end{equation}
where the quasiparticle weight matrix is given by
\begin{equation} \label{eq:Z}
\bm{Z} \equiv \left( \bm{1} - \left. \frac{\partial \mathrm{Re} \bm{\Sigma}(\omega) }{\partial \omega} \right|_{\omega = 0} \right)^{-1} = \bm{\mathcal{R}}^{\dagger} \bm{\mathcal{R}}^{}.
\end{equation}
It is easily checked that the quantities of the physical electron $\bm{G}(k,\omega)$, $\bm{\Sigma}(\omega)$, $\bm{A}(k,\omega)$, and $\bm{Z}$ are gauge invariant \cite{Lechermann2007}. We present the results below in the trimer basis of a triangle (\cref{fig:trimer-ground-state}a), which elucidates the connection to the ground state of an isolated triangle and the physical mechanism of the spin-$1$ Mott insulator.

We implemented RISB within the \verb+TRIQS+ library \cite{Parcollet2015,Seth2016}. The $k$-integrals were evaluated using the linear tetrahedron method \cite{Blochl1994}, and the impurity was solved using exact diagonalization using the Arnoldi method in \verb+ARPACK-NG+ \cite{TRIQS-ARPACK}. On the embedding state $|\Phi\rangle$ and mean-field matrices $\bm{\lambda}$, $\bm{\mathcal{R}}$, $\bm{\lambda}^c$, and $\bm{\Delta}^p$, we imposed the symmetry of $\hat{H}_i^{\triangle}$ including the $\mathcal{C}_3$ rotational symmetry of a triangle, but we relaxed the full SU($2$) symmetry when investigating antiferromagnetic solutions. The matrices are block diagonal in spin, and in the trimer basis are block diagonal in the trimer orbitals $A$, $E_1$, and $E_2$, where the matrix elements for $E_1$ and $E_2$ are equivalent because of the $\mathcal{C}_3$ symmetry (and we denote either by $E$).

\section{Spin-$1$ Mott metal-insulator transition: paramagnetic solutions} \label{sec:para}

We first investigate paramagnetic solutions obtained within RISB. We will show that there is a Mott metal-insulator transition to a spin-$1$ state with a vanishing quasiparticle weight on the decorated honeycomb lattice. In the spin-$1$ Mott insulator electrons become localized to a triangle and within RISB charge fluctuations between triangles are frozen. The spin-$1$ Mott insulator is adiabatically connected to the limit of isolated triangles.

This is the usual description of a Mott insulator captured in slave boson theories \cite{Kotliar1986}, and originally described by Brinkman and Rice for the half-filled Hubbard model \cite{Brinkman1970}. We highlight that our results will show that the spin-$1$ Mott metal-insulator transition is first order (\cref{fig:para-hopping-trimer}), which differs from the Brinkman-Rice mechanism which is continuous. Furthermore, the description within RISB is richer than a simpler single-site theory because correlations renormalize the quasiparticle bands by different amounts, which is essential to capture the spin-$1$ Mott insulator.

The spin-$1$ Mott insulator occurs for a wide range of $J_{\mathrm{GK}}$ (\cref{fig:para-effective-J}). We provide detailed results for $J_{\mathrm{GK}} = 0$, but qualitatively similar results are found for $J_{\mathrm{GK}} \le 0$. This corresponds to the single-orbital Hubbard model on the decorated honeycomb lattice.

\subsection{Renormalized hopping amplitudes} \label{sec:renorm-hopping}

A useful way to interpret results within RISB is with the auxiliary fermions of the theory, which are described by the Hamiltonian $\hat{H}^{\mathrm{qp}}$ (\cref{eq:ham-qp}). Within this picture RISB maps the interacting system to a non-interacting Hamiltonian with parameters renormalized due to the weights of local electronic configurations on a triangle. 

At $U=0$ $\hat{H}^{\mathrm{qp}}$ coincides with the physical Hamiltonian (\cref{eq:ham}) because $\bm{\mathcal{R}} = \bm{1}$ and $\bm{\lambda} = \bm{\varepsilon}^0$, where $\bm{\varepsilon}^0$ are the one-body terms in $\hat{H}^{\triangle}$. When interactions are turned on the parameters are renormalized. The renormalized inter-trimer hopping terms are given by the matrix $\bm{t}_{ij}^* \equiv \bm{\mathcal{R}}^{} \bm{t}_{ij} \bm{\mathcal{R}}^{\dagger}$, and the renormalized intra-trimer terms are captured in the matrix $\bm{\lambda}$. 

The matrices $\bm{\mathcal{R}}$ and $\bm{\lambda}$ are block diagonal in the $A$ and $E$ orbital representation of a trimer for states that do not break the $\mathcal{C}_3$ rotational symmetry of a triangle. The correlation potential matrix $\bm{\lambda}$ captures the energy difference between the $A$ and degenerate $E$ trimer orbitals, given by $[\bm{\lambda}]_{E} - [\bm{\lambda}]_{A}$. Hence, an auxiliary fermion can only hop between the $A$ and $E$ orbitals by hopping to an adjacent triangle with amplitude $[\bm{t}_{ij}^*]_{AE} = [\bm{t}_{ij}^*]_{EA}$.

\begin{figure}
	\centering
	\includegraphics[width=\columnwidth]{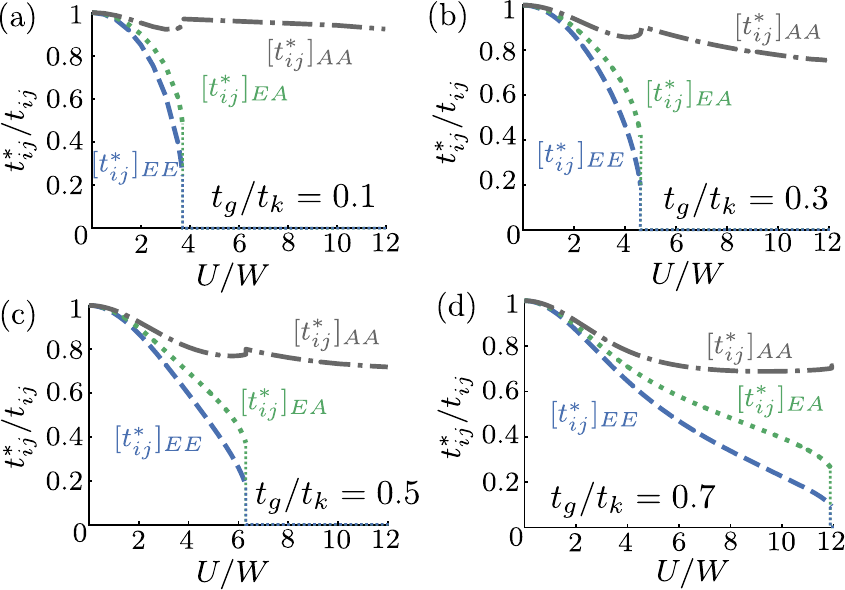}
	\caption{\label{fig:para-hopping-trimer}
		Metal-insulator transition. Renormalized inter-triangle hopping amplitudes in the trimer basis in the paramagnetic state, given by $[\bm{t}_{ij}^*]_{mn} = [\bm{\mathcal{R}}_i^{} \bm{t}_{ij} \bm{\mathcal{R}}_j^{\dagger}]_{mn}$, where $m,n = A,E$. In the insulator hopping between the $E$ trimer orbitals of adjacent triangles vanishes while hopping between the $A$ orbitals remains finite. The non-interacting bandwidth is $W = 2t_g$.
	}
\end{figure}

\begin{figure}
	\centering
	\includegraphics[width=\columnwidth]{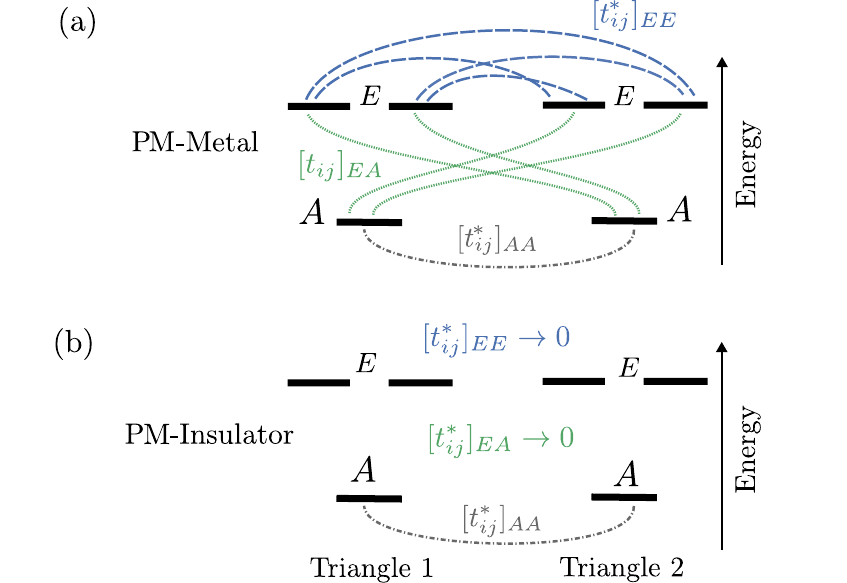}
	\caption{\label{fig:para-hopping-schematic}
		Schematic of the renormalized inter-triangle hopping parameters in the trimer basis in the paramagnetic (PM) state. (a) The hopping parameters remain finite between all trimer orbitals of adjacent triangles in the metal. (b) Inter-triangle hopping involving the $E$ orbitals of a trimer vanishes in the insulating phase.
	}
\end{figure}

There is a metal-insulator transition signaled by vanishing hopping between triangles. At the transition $[\bm{\mathcal{R}}]_{EE} \rightarrow 0$, resulting in $[\bm{t}_{ij}^*]_{EE} = [\bm{t}_{ij}^*]_{AE} = [\bm{t}_{ij}^*]_{EA} = 0$, while $[\bm{t}_{ij}^*]_{AA} \neq 0$ (\cref{fig:para-hopping-trimer}). The $A$ and $E$ orbitals decouple (\cref{fig:para-hopping-schematic}). Even though $[\bm{t}_{ij}^*]_{AA} \neq 0$, the lower set of bands in the insulator have only $A$ orbital character and are fully occupied at two-thirds filling. Hence, in the auxiliary fermion system there is no hopping between clusters and there are four auxiliary fermions localized to each triangle.

\begin{figure}
	\centering
	\includegraphics[width=\columnwidth]{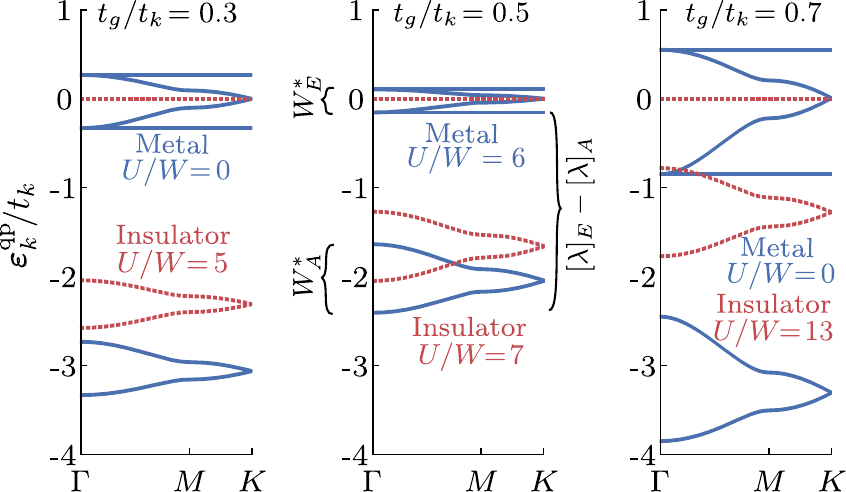}
	\caption{\label{fig:para-dipersion}
	The renormalized bandwidth $W_E^* \equiv 2 [\bm{t}_{ij}^*]_{EE}$ associated with the set of $E$ bands narrows as correlations increase and are flat in the insulator, while the renormalized bandwidth $W_A^* \equiv 2 [\bm{t}_{ij}^*]_{AA}$ associated with the $A$ bands remains finite and fully occupied. The dispersion $\bm{\varepsilon}_k^{\mathrm{qp}}$ are the eigenenergies of the auxiliary fermion Hamiltonian \cref{eq:ham-qp}.
	}
\end{figure}

The vanishing hopping between triangles results in flat bands at the Fermi energy in the spectrum of the auxiliary fermions. The $E$ and $A$ bands (\cref{sec:model}) have bandwidths $W_E^* \equiv 2[\bm{t}_{ij}^*]_{EE}$ and $W_A^* \equiv 2[\bm{t}_{ij}^*]_{AA}$ respectively, and are separated in energy by $[\bm{\lambda}]_{E} - [\bm{\lambda}]_{A}$. As the insulator is approached from the metallic side the bands narrow until $W_E^* \rightarrow 0$ at the transition, while $W_{A}^*$ has finite bandwidth but is fully occupied (\cref{fig:para-dipersion}). Hence, the spectrum of the auxiliary fermions has flat bands at the Fermi energy, signifying that the state is insulating.

\subsection{Quasiparticle weight and quasiparticle bands} \label{sec:quasiparticle-bands}

While the auxiliary fermions provide insight into the metal-insulator transition, the spectrum of $\hat{H}^{\mathrm{qp}}$ is not equivalent to the spectrum of the physical electron. Instead, the real spectrum is made up of many-body excitations given by the unitary transformation \cref{eq:sb-electron} and captured by the renormalization matrix $\bm{\mathcal{R}}$. Within RISB the result is a quasiparticle description, whose propagator is given by the physical electron Green's function (\cref{eq:greens-fnc}) and is entirely coherent.

One can understand the effect of correlations on electrons by considering the spectral function (\cref{eq:spectral-function}). Within RISB the coherent part of the single-particle spectral function, as measured in angle-resolved photoemission spectroscopy (ARPES), is given by
\begin{align} \label{eq:arpes-spectral-function}
A(k,\omega) & \equiv -\pi^{-1} \mathrm{Im} \mathrm{Tr} \bm{G}(k,\omega) \nonumber \\
& = \mathrm{Tr} ( \bm{\mathcal{R}}^{\dagger} \delta(\omega \bm{1} - \bm{\varepsilon}_k^{\mathrm{qp}}) \bm{\mathcal{R}}^{}) \nonumber \\
& = \sum_{p} Z_p^{\mathrm{qp}}(k) \delta(\omega - \varepsilon_{kp}^{\mathrm{qp}}),
\end{align}
where the quasiparticle weight in band $p$ is
\begin{equation} \label{eq:Z-qp}
Z_p^{\mathrm{qp}}(k) \equiv [\bm{U}^{\dagger}(k) \bm{Z} \bm{U}(k)]_{pp},
\end{equation}
with the spectral weight in the orbitals (\cref{eq:spectral-weight}) given by $\bm{Z} = \bm{\mathcal{R}}^{} \bm{\mathcal{R}}^{\dagger} = \bm{\mathcal{R}}^{\dagger} \bm{\mathcal{R}}$ because $\bm{\mathcal{R}}$ is unitary, and $\bm{U}(k)$ is the single-particle unitary transformation that diagonalizes $\hat{H}^{\mathrm{qp}}$ (\cref{eq:ham-qp}) with eigenenergies $\varepsilon_{kp}^{\mathrm{qp}}$. $Z_p^{\mathrm{qp}}(k) \in [0,1]$, with $Z_p^{\mathrm{qp}}(k) = 1$ corresponding to the non-interacting problem ($U=0$), and increasing correlations generically decreases $Z_p^{\mathrm{qp}}(k)$.

At two-thirds filling the quasiparticle weight $[\bm{Z}]_E$ in the $E$ orbitals of a trimer approximates the quasiparticle weight of the bands near the Fermi energy \cite{Nourse2021}. Hence, $[\bm{Z}]_E$ is a measure of the metallicity of the system, with $[\bm{Z}]_E > 0$ corresponding to a metal and $[\bm{Z}]_E = 0$ to a correlated insulator. A vanishing quasiparticle weight signifies a breakdown of the Fermi liquid with a vanishing quasiparticle peak at the Fermi energy in the single-particle spectral function $A(\omega)$ of the electrons (\cref{eq:arpes-spectral-function}).

\begin{figure}
	\centering
	\includegraphics[width=\columnwidth]{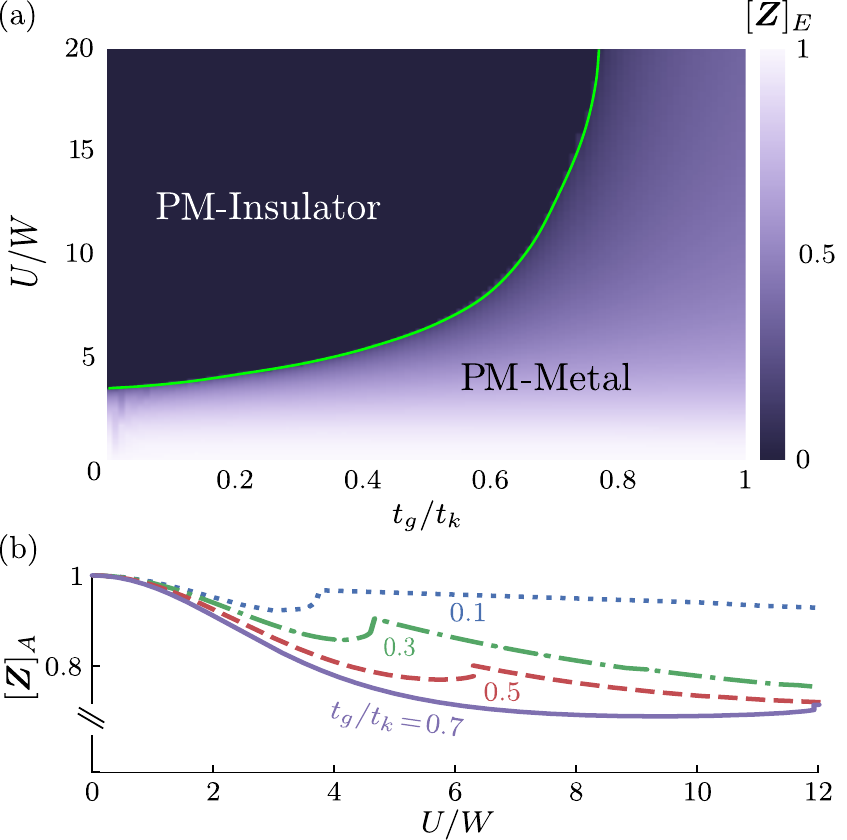}
	\caption{\label{fig:Z}
		The metal-insulator transition can be driven by correlations ($U/W$) or moving towards the molecular limit ($t_g/t_k \rightarrow 0$), as signified by vanishing elements of the quasiparticle weight matrix $\bm{Z}$ in the trimer basis.
		(a) In the paramagnetic (PM) insulator the quasiparticle weight associated with the $E$ molecular orbitals $[\bm{Z}]_E$ discontinuously vanishes. The solid green line marks the phase boundary.
		(b) $[\bm{Z}]_A$ does not vanish. However, the $A$ molecular orbitals are fully occupied in the insulator (cf. \cref{fig:para-orbital-fluct}a).
		There is no Mott insulator for $t_g / t_k \gtrsim 0.86$. 
	}
\end{figure}

In \cref{fig:Z} we show the quasiparticle weight in the trimer orbitals for different hopping ratios $t_g/t_k$. The transition from a Dirac metal to a Mott insulator is indicated by a vanishing $[\bm{Z}]_E$ (\cref{fig:Z}a)]. Even though $[\bm{Z}]_A$ remains finite in the insulator (\cref{fig:Z}b) it characterizes the spectral weight in quasiparticle bands away from the Fermi energy that are fully occupied. For $t_g/t_k \gtrsim 0.86$ there is no insulator for finite $U$ and there instead is a correlated metal. The insulator is driven either by electronic correlations, or moving towards the molecular limit by lowering $t_g/t_k$. 

\begin{figure}
	\centering
	\includegraphics[width=\columnwidth]{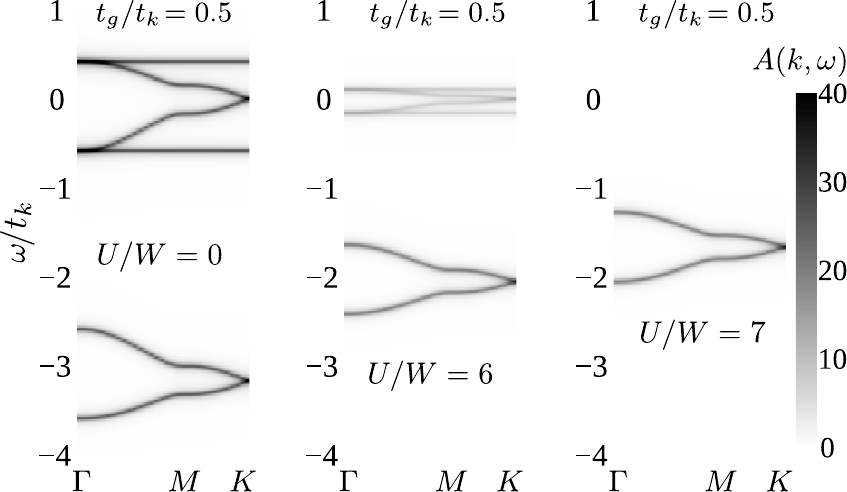}
	\caption{\label{fig:para-spectral-fnc}
	Spectral weight vanishes in the quasiparticle bands near the Fermi energy ($\omega = 0$) at the metal-insulator transition. The bands are given by the spectrum of the auxiliary fermions (cf.~\cref{fig:para-dipersion}) with spectral weight $A(k,\omega)$ renormalized by the quasiparticle weight. A broadening factor of $+i\eta = i0.025t_k$ was used because the self-energy $\bm{\Sigma}(\omega)$ is real in the mean-field RISB approximation.
	}
\end{figure}

In \cref{fig:para-spectral-fnc} we show the corresponding spectral function $A(k,\omega)$ as the metal-insulator transition is approached. The quasiparticle bands are given by the spectrum of $\hat{H}^{\mathrm{qp}}$ (cf.~\cref{sec:renorm-hopping}): $\bm{Z}$ narrows the bandwidth by renormalizing the inter-cluster hopping and $\bm{\lambda}$ shifts the position of the bands. The spectral weight in each band is reduced because of the renormalization by $\bm{Z}$ and in the insulator vanishes for the bands at the Fermi energy. 

\subsection{Spin-1 formation on a triangle}

\begin{figure}
	\centering
	\includegraphics[width=\columnwidth]{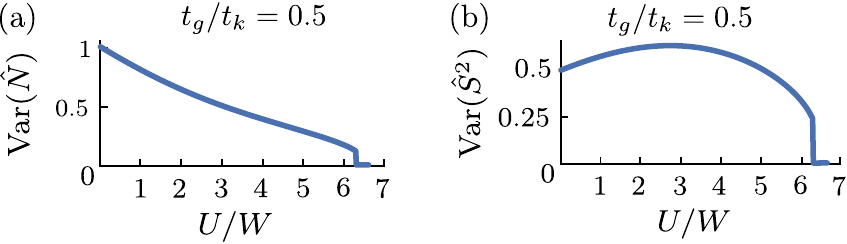}
	\caption{\label{fig:para-triangle-fluct}
		(a) Charge and (b) spin fluctuations on a triangle vanish in the paramagnetic insulator because electrons become localized within a triangle. The variance of a generic local operator $\hat{X}$ is $\mathrm{Var}(\hat{X}) = \sum_i (\hat{X}_i - \langle \hat{X}_i \rangle)^2 / 2 \mathcal{N}$, where $\hat{X}_i = \sum_{\alpha=1}^3 \hat{X}_{i\alpha}$, and $\mathcal{N}$ is the number of unit cells on the lattice. The total particle number operator of triangle $i$ is $\hat{N}_i = \sum_{\alpha=1}^3 \sum_{\sigma} \hat{n}_{i\alpha\sigma}$.
	}
\end{figure}

We now wish to characterize the Mott insulating phase by looking at the local properties on a triangle. As $U$ increases the charge and spin fluctuations on a triangle are suppressed, vanishing at the metal-insulator transition (\cref{fig:para-triangle-fluct}). In the Mott insulator, electrons become localized to a triangle with four electrons per triangle. Charge and spin fluctuations between triangles are missed by the RISB approximation.

\begin{figure}
	\centering
	\includegraphics[width=\columnwidth]{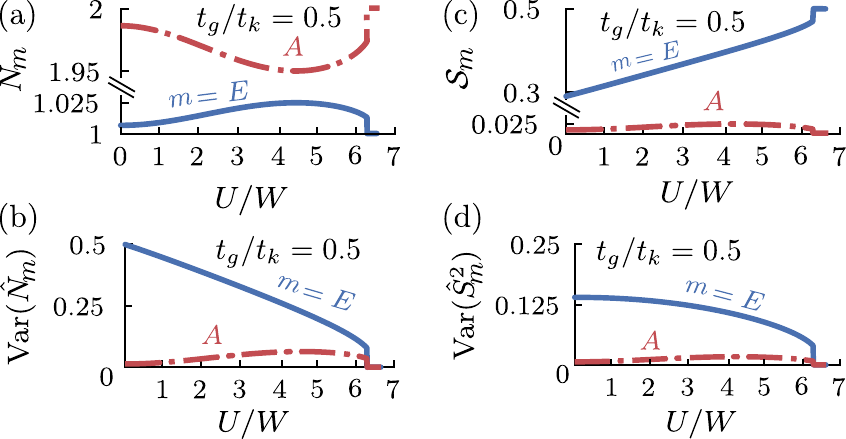}
	\caption{\label{fig:para-orbital-fluct}
		(a) Number of particles in orbital $m$ of a trimer. In the insulator the $A$ orbital is fully occupied and there are two electrons shared between the $E$ orbitals.
		(b) Charge fluctuations of a trimer orbital vanish in the insulator in the RISB approximation.
		(c) Total spin in the trimer orbitals. In the insulator the electrons localized to the $E$ orbitals are pseudo-spin-1/2 moments, and combine to form spin-triplets (cf.~\cref{fig:spin-moment}).
		(d) Spin fluctuations of a trimer orbital vanish in the insulator in the RISB approximation.
	}
\end{figure}

\begin{figure}
	\centering
	\includegraphics[width=\columnwidth]{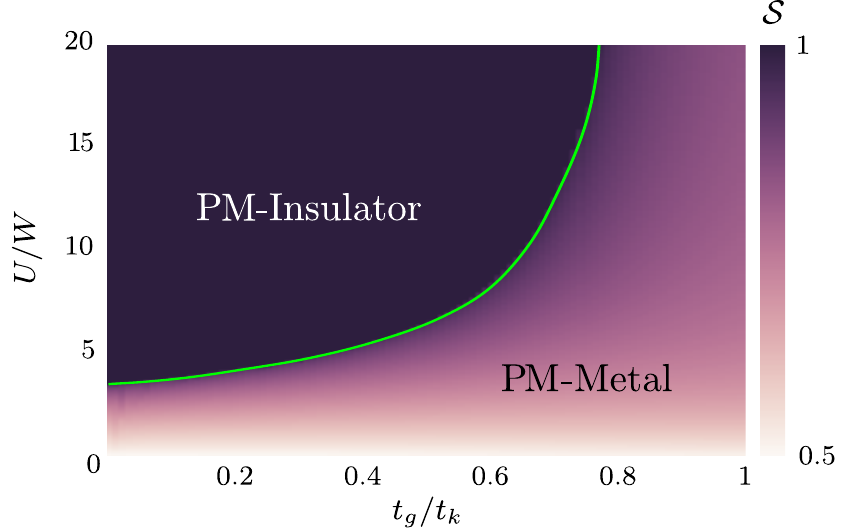}
	\caption{\label{fig:spin-moment}
	Increasing interactions favors spin-triplet electronic configurations on a triangle. Spin-$1$ moments form in the Mott insulator. The solid green line marks the phase boundary.
	}
\end{figure}

In the trimer basis, the effective Coulomb repulsion $U/3$ suppresses intra-orbital charge and spin fluctuations (\cref{fig:para-orbital-fluct}b,d). Electrons become localized to the $A$ and $E$ orbitals in the insulator, with the $A$ orbitals fully occupied and two electrons shared by the degenerate $E$ orbitals (\cref{fig:para-orbital-fluct}a).

The electrons localized to the $E$ orbitals act as effective spin-$1/2$ degrees of freedom (\cref{fig:para-orbital-fluct}c), and combine to form a total spin-triplet on each triangle (\cref{fig:spin-moment}). When there are four electrons per triangle $\mathcal{S} = 0$ indicates only spin-singlets while $\mathcal{S} = 1$ indicates spin-triplet electronic configurations. Hence, we denote the insulator as a spin-$1$ Mott insulator.

RISB captures the insulator as the molecular limit ($t_g / t_k \rightarrow 0$) of \cref{eq:ham}, similarly to how Kotliar-Ruckenstein slave bosons capture a Mott insulator at half-filling as the atomic limit with no spin fluctuations. One can see this by looking at the parameters of the embedding Hamiltonian (\cref{eq:ham-emb}) in the insulator. Within RISB the energy per triangle is given by
\begin{equation} \label{eq:energy}
	E \equiv \sum_{a\alpha} [\bm{\mathcal{D}}]_{a\alpha} \langle \Phi | \hat{c}_{\alpha}^{\dagger} \hat{f}_{a}^{} | \Phi \rangle + \langle \Phi | \hat{H}^{\triangle} | \Phi \rangle.
\end{equation}
In the insulator all elements of $\bm{\mathcal{D}}$ are numerically zero so that the impurity decouples from the bath. The impurity state $|\Phi \rangle$ of the insulator is a spin-triplet on each triangle with four electrons and is an eigenstate of $\hat{H}^{\triangle}$. Hence, the ground state energy per triangle in the paramagnetic insulator is equivalent to the ground state energy of an isolated triangle, given by \cref{eq:energy-isolated-triangle}.

Therefore, the spin-$1$ Mott insulator is an extended phase from the ground state of an isolated triangle where inter-triangle coupling is not sufficient to destroy spin-triplet formation arising from strong correlations. Because the spin-$1$ Mott insulator realizes the molecular limit, a minimal model of the metal-insulator transition may be captured by a degenerate two-orbital Hubbard-Kanamori model (\cref{sec:non-interacting-limit}).

\section{Spin-state transitions from intra-triangle spin exchange} \label{sec:spin-state-transitions}

In this section we investigate the effect of intra-triangle spin exchange $J_{\mathrm{GK}}$ (\cref{eq:ham-F}). We will show that $J_{\mathrm{GK}}$ can drive a spin-state transition from a spin-$1$ Mott insulator via a metallic phase to a spin-$0$ Mott insulator for an antiferromagnetic coupling ($J_{\mathrm{GK}} > 0$). This highlights that the spin-$1$ Mott insulator is driven by an effective Hund's rule coupling that favors spin-triplet formation on a triangle. The spin-state transition can be simply understood from the energetics in the limit of isolated triangles.

We only investigate an intra-triangle spin exchange, which is simple to include with the local interaction within a triangle ($\hat{H}^{\triangle}$ in \cref{eq:ham}). This is justified because for most materials in the small $t_g/t_k$ regime intra-triangle spin exchange dominates over the inter-triangle spin exchange. With our choice of triangular clusters, an inter-triangle spin exchange can only be treated at the Hartree-Fock level. There are certainly systems where this approximation is not justified (\cref{sec:gka-rules}). However, there are many materials where our model should be appropriate.

To elucidate the effect of $J_{\mathrm{GK}}$ it is useful to define $J \equiv U + 2 J_{\mathrm{GK}}$ and $r \equiv 3 J_{\mathrm{GK}}/ J$, so that $r$ controls the relative strength of the total spin $\vec{S}_i^2$ term. The local interactions on a triangle (\cref{eq-ham-U-trimer,eq-ham-F-trimer}) can be written as
\begin{align} \label{eq:ham-hunds}
& \hat{H}_i^U  + \hat{H}_i^J  = - \frac{J}{3} \vec{S}_i^2(1 - r) + \frac{J}{12}(1 - r )  \hat{N}_i^2 \nonumber \\
&+ \frac{J}{6} \left( 1 - r \right) \hat{N}_i - \frac{J}{3} \sum_m \hat{n}_{im\uparrow} \hat{n}_{im\downarrow} \nonumber \\
& + \frac{J}{3} \sum_{m} \sum_{n \neq m} \sum_{p \neq m \neq n} \left( \hat{b}_{im \downarrow}^{\dagger} \hat{b}_{im \uparrow}^{\dagger} \hat{b}_{in \uparrow}^{} \hat{b}_{ip \downarrow}^{} + \mathrm{H.c.} \right).
\end{align}
$r$ controls the relative strength of the $\vec{S}_i^2$ term. Hence, when there are four electrons on a triangle, $J_{\mathrm{GK}}$ determines whether spin singlets or spin triplets are favored. We also highlight that at $U = - 2 J_{\mathrm{GK}}$ \cref{eq:ham-hunds} is equivalent to the Hubbard-Kanamori Hamiltonian, up to a constant, for the parameters $U' = U - J$, $U = 19 J / 2$, $J_P = 0$, and $J_X = J$ (see Eq.~5 in \cite{Georges2013}).

\begin{figure}
	\centering
	\includegraphics[width=\columnwidth]{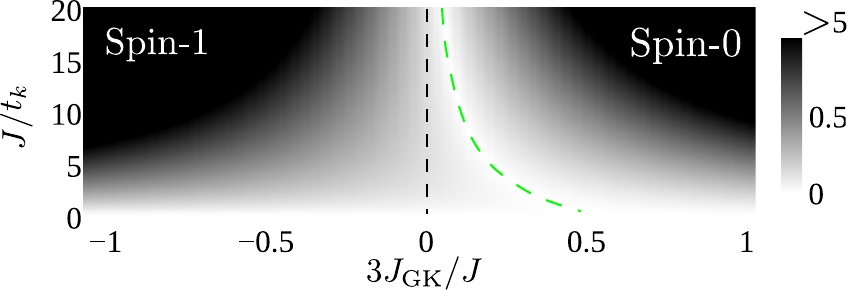} 
	\caption{\label{fig:isolated-effective-J}
		Energy difference between the spin singlet and spin triplet electron configuration on an isolated triangle at two-thirds filling. $J_{\mathrm{GK}}$ tunes between a triplet ground state and a singlet ground state. The green dashed line denotes the boundary between the spin-$0$ and spin-$1$ ground states, the grayscale is the absolute energy difference between the lowest energy spin-$0$ and lowest energy spin-$1$ states, and the black dashed line is the Hubbard model where $J_{\mathrm{GK}} = 0$.
	}
\end{figure}

One can gain insight into the overall effect of $J_{\mathrm{GK}}$ by considering an isolated triangle. In \cref{fig:isolated-effective-J} we plot the energy difference between the spin-singlet and spin-triplet states, which identifies the ground state. Roughly, for large $J$ a spin-triplet (spin-singlet) ground state is favored for $r \le 0$ ($r > 0$). For $J$ small enough a spin-triplet is favored even for $r < 1/2$, but one can check numerically that the boundary between the spin-singlet and spin-triplet approaches $r = 0$ as $J \rightarrow \infty$.

\begin{figure}
\centering
\includegraphics[width=\columnwidth]{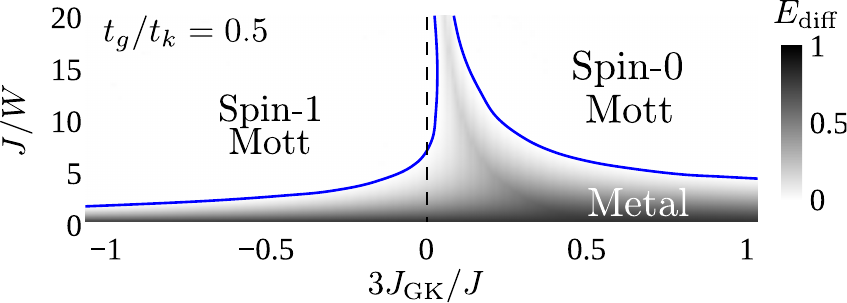}
\caption{\label{fig:para-effective-J-energy}
	Energy per triangle $E$ compared to the ground state energy of an isolated triangle $E_0^{\triangle}$, $E_{\mathrm{diff}} = E - E_0^{\triangle}$. The insulators within RISB realize the molecular limit. Hence, the spin-$1$ and spin-$0$ Mott insulators are adiabatically connected to the ground state of an isolated triangle (cf.~\cref{fig:isolated-effective-J}).
	The dashed black line marks $J_{\mathrm{GK}} = 0$, and the solid blue lines mark the phase boundaries.
}
\end{figure}

In \cref{fig:para-effective-J} we show the effect of varying $r$ on the decorated honeycomb lattice for $t_g/t_k = 0.5$. For $r < 0 $ the $\vec{S}_i^2$ term in \cref{eq:ham-hunds} reflects Hund's first rule which drives spin-triplet formation on a triangle and lowers the critical interaction strength of the metal-insulator transition. This effect is similarly observed in Hund's metals at half-filling \cite{Lechermann2007, Georges2013}. The insulator at $r < 0$ is in the same phase as the spin-$1$ Mott insulator at $r = 0$. For $r > 0$ an insulator still occurs for sufficiently large $J$, but each triangle forms a total spin-$0$ configuration. Both insulators are adiabatically connected to the ground state of an isolated triangle (\cref{fig:para-effective-J-energy}). In the small region for $r > 0$ where the spin-$1$ Mott insulator occurs there is an insulator-metal transition at larger $J/W$, which follows the green dashed line in \cref{fig:isolated-effective-J}.

Hence, it is possible to tune the system between a spin-$1$ Mott insulator, a metal, and a spin-$0$ Mott insulator. Below we outline how $J_{\mathrm{GK}}$ may be estimated and tuned in materials where decorated lattices commonly occur.

It is worth comparing the spin-$0$ Mott insulator on the decorated honeycomb lattice to other spin-$0$ Mott insulators in multi-orbital Hubbard models. In some cases a spin-$0$ Mott insulator is adiabatically connected to a band insulator, such as the crossover from a band insulator to a $t_g$-dimer valence bond solid \cite{Nourse2021}, and the Mott-Peierls crossover in the dimer lattice model \cite{Fabrizio2007,Najera2018}. It is unclear how, or if, the spin-$0$ Mott insulator in this paper adiabatically connects to a band insulator without explicitly breaking a symmetry. It has been shown that there exist so-called fragile Mott insulators \cite{Yao2010}. A fragile Mott insulator transforms under a non-trivial one-dimensional representation of the crystal point group, while a band insulator transforms under the trivial representation. The spin-$0$ Mott insulator in this paper is in the trivial representation and is therefore not a fragile Mott insulator. However, this does not necessarily imply that it is adiabatically connected to a band insulator.

\subsection{Estimating $J_{\mathrm{GK}}$ in coordination polymers} \label{sec:gka-rules}
In coordination polymers it is thought that the dominant spin-exchange process between metals occurs from kinetic exchange through a bridging ligand, with a M-L-M configuration \cite{Thorarinsdottir2020}. There are other processes that contribute to exchange, such as a ferromagnetic two-electron exchange between adjacent metals, but they are thought to be small and are often neglected \cite{Goodenough1963,Weihe2000,Kenny2021}.

In simple systems composed of metals bridged by a single-atom ligand the Goodenough-Kanamori-Anderson (GKA) rules are often used to estimate the exchange \cite{Goodenough1955,Goodenough1958,Goodenough1963,Kanamori1959,Kanamori1963b,Anderson1959,Anderson1963}. The symmetry relationship between the orbitals, their occupancy, and the angle of the M-L-M bond determines the magnitude of the exchange and whether it is antiferromagnetic or ferromagnetic. For these simple systems more quantitative estimates have been made, using Anderson's original approach \cite{Anderson1963,Geertsma1990,Weihe1997}, with valence bond configuration interaction methods \cite{Zaanen1987,Tuczek1993,Brown1995,Weihe2000}, as well as other proposals for when some of the assumptions are not met \cite{Heuvel2007}. However, the application of the GKA rules do not always work. Multiple competing exchange interactions, deviations in bond angles, the type of metal-ligand bond, and the presence of multi-atom bridges makes predicting the sign and magnitude of the exchange difficult, as is found in some magnetic organometallics \cite{Geertsma1995,Geertsma1996} and metal-organic framework magnets \cite{Thorarinsdottir2020}. Furthermore, spin-orbital entanglement may violate the GKA rules \cite{Oleifmmode2006}. Regardless, the GKA rules often correctly predict the sign of the exchange in coordination polymers \cite{Thorarinsdottir2020}.

The GKA rules for superexchange are the following:
\renewcommand{\labelenumi}{\roman{enumi}.}
\begin{enumerate} [nosep]
	\item strong antiferromagnetic coupling between two metals that occurs from half-filled symmetry-compatible orbitals (typically d orbitals where the M-L-M bond is $\sim \! 180^{\circ}$);
	\item weak ferromagnetic coupling between two metals when there are half-filled symmetry-incompatible orbitals (typically d orbitals where the M-L-M bond is $\sim \! 90^{\circ}$);
	\item weak ferromagnetic coupling between two metals when one orbital is half-filled and the other is empty or fully occupied.
\end{enumerate}
There is an additional contribution from an empty orbital on a metal and a fully occupied orbital on an adjacent metal that can be ferromagnetic \cite{Anderson1963,Weihe2000} or antiferromagnetic \cite{Weihe1997}, but it tends to be neglected because it is much weaker than the other exchange processes.

Many metal-organic framework magnets have multi-atom ligands that are $\pi$-conjugated. This provides an additional pathway for spin exchange because an unpaired electron on one atom polarizes the electrons on adjacent atoms, which overall favors antiparallel spin arrangements \cite{Thorarinsdottir2020}. The overall spin exchange between adjacent metals from this process is (anti-)ferromagnetic when the magnetic $\pi$-pathway through the bridging ligand has an (even) odd number of atoms \cite{Thorarinsdottir2020}. Through space interactions are also important in some coordination polymers \cite{Kenny2021}.

$J_{\mathrm{GK}}$ can be tuned with physical or chemical pressure, which changes the angle of the M-L-M bond, as observed in copper oxides \cite{Mizuno1998,Shimizu2003}. It has been shown theoretically that pressure is enough to change the sign of $J_{\mathrm{GK}}$ \cite{Rocquefelte2012}.

\section{Spin-$1$ Slater metal-insulator transition: antiferromagnetic solutions} \label{sec:afm}

We now describe the magnetically ordered states captured within RISB. We will show that there is a Slater insulator \cite{Slater1951} caused by a spin-density wave (SDW) between adjacent spin-$1$ polarized triangles which opens a gap in the quasiparticle spectrum. The Slater insulator occurs at a smaller critical interaction strength than the Mott insulator presented in \cref{sec:para}. The ordered state occurs because of an energy cost for double occupancy so that it becomes energetically favorable to arrange each spin on separate sublattices $\mathcal{A}$ and $\mathcal{B}$. A SDW lowers the overall energy compared to a paramagnetic metal by lowering the energy of the highest occupied quasiparticle band.

To capture the antiferromagnetic order we allow the set of inequivalent triangles $\mathcal{A}$ and $\mathcal{B}$ (see \cref{fig:lattice}a) to be described by a different embedding state $|\Phi_{\ell}\rangle$ (see \cref{eq:selfcon-4}) where $\ell = \mathcal{A}, \mathcal{B}$. However, only a single embedding state has to be solved because the magnetic order between triangles at the mean-field level can be accounted for by using the unitary transformation
\begin{equation}
\bm{U}_{\ell} = e^{i\phi_{\ell} \bm{\tau}^y/2},
\end{equation}
where $\bm{\phi}_{\ell} = 0$ ($\pi$) for $\ell = \mathcal{A}$ ($\mathcal{B}$), and $\bm{\tau}^y$ is the Pauli matrix. The transformation can be described by a $2\times2$ matrix which acts on the spin degrees of freedom and transforms the mean-field matrices as
\begin{align}
[\bm{\lambda}_{\mathcal{B}}]_{m} & = \bm{U}_{\mathcal{B}}^{\dagger} [\bm{\lambda}_{\mathcal{A}}]_{m} \bm{U}_{\mathcal{B}}^{}, \nonumber \\
[\bm{\mathcal{R}}_{\mathcal{B}}]_{m} & = \bm{U}_{\mathcal{B}}^{\dagger} [\bm{\mathcal{R}}_{\mathcal{A}}]_{m} \bm{U}_{\mathcal{B}}^{}.
\end{align}

\begin{figure}
	\centering
	\includegraphics[width=\columnwidth]{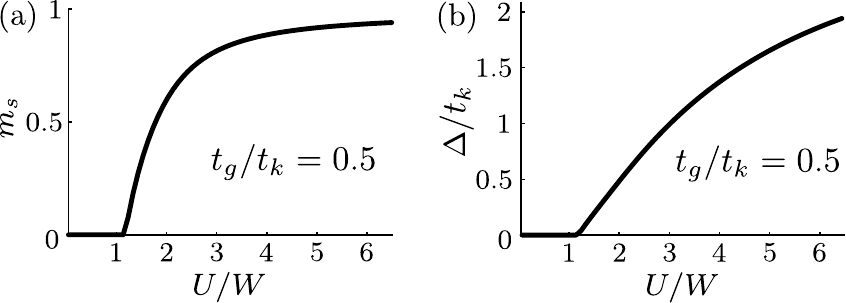}
	\caption{\label{fig:afm-order-parameter}
		There is an instability towards antiferromagnetism, signified by the
		(a) staggered order parameter $m_s$ between adjacent triangles and a
		(b) gap $\Delta$ opening in the quasiparticle bands (cf.~\cref{fig:afm-dispersion}).
	}
\end{figure}

\begin{figure}
	\centering
	\includegraphics[width=\columnwidth]{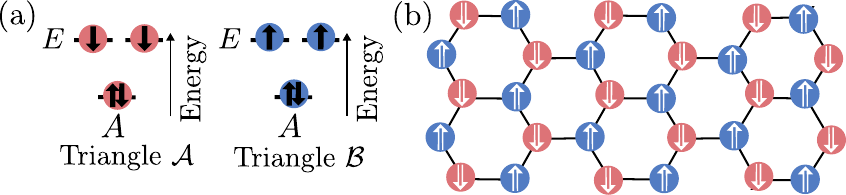}
	\caption{\label{fig:afm-neel-order}
		Schematic of the spin-$1$ antiferromagnetic ground state.
		(a) Spin polarized spin-triplet configurations form in the antiferromagnetic insulator
		and (b) alternate on each triangle.
	}
\end{figure}

In \cref{fig:afm-order-parameter}a we show the staggered order parameter, given by \begin{equation}
	m_s \equiv \frac{1}{2 \mathcal{N}} \left| \sum_{i} (-1)^{\eta(i)} \langle \hat{S}_i^{z} \rangle \right|,
\end{equation}
where $\mathcal{N}$ is the number of unit-cells, and $\eta(i) = 0$ ($1$) if the triangle is on sublattice $\mathcal{A}$ ($\mathcal{B}$). A finite $m_s$ signifies the formation of a SDW on the decorated honeycomb lattice (\cref{fig:afm-neel-order}) due to spin polarization on a triangle. When $m_s \rightarrow 1$ the triangles are fully polarized and there are only $S_z = \pm 1$ spin-triplet electronic configurations. At the onset of antiferromagnetism a gap $\Delta$ opens continuously at the Fermi energy (\cref{fig:afm-order-parameter}(b)), where $\Delta$ is the energy difference between the lowest unoccupied and highest occupied quasiparticle band.

\begin{figure}
	\centering
	\includegraphics[width=\columnwidth]{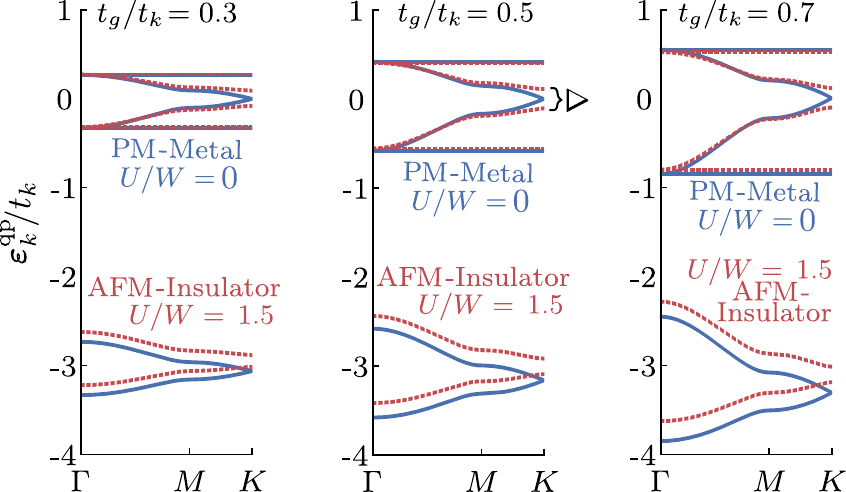}
	\caption{\label{fig:afm-dispersion}
		Band structures of the auxiliary fermions in the antiferromagnetic (AFM) state. In the paramagnetic (PM) Dirac metal the Fermi energy is at the high symmetry point $K$. In the insulator the inversion symmetry about the center of the bond between adjacent triangles is broken and a gap $\Delta$ opens (cf.~\cref{fig:afm-order-parameter}b). The gap $\Delta$ is given by the energy difference between the highest occupied band and the lowest unoccupied band.
	}
\end{figure}

\begin{figure}
	\centering
	\includegraphics[width=\columnwidth]{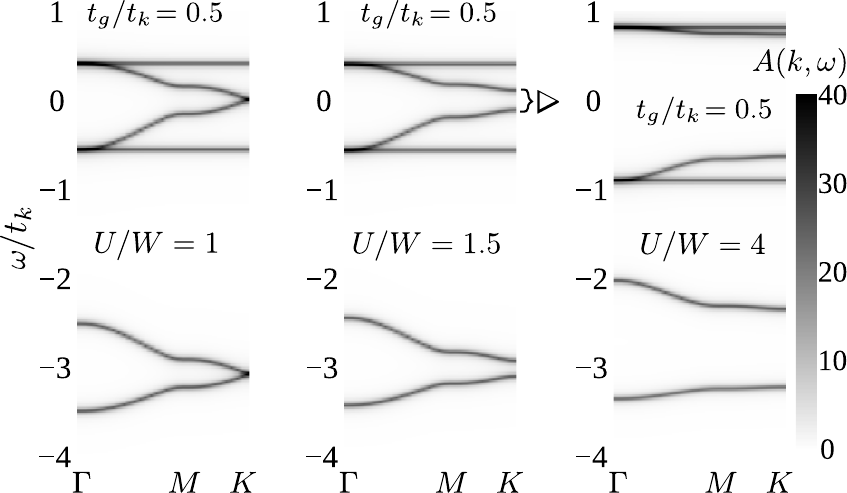}
	\caption{\label{fig:afm-spectral-fnc}
		Quasiparticle bands in the antiferromagnetic state. The bands occur at the spectrum of the auxiliary fermions (cf.~\cref{fig:afm-dispersion}) with spectral weight $A(k,\omega)$ renormalized by the quasiparticle weight. In the antiferromagnetic insulator a gap $\Delta$ opens at the high symmetry point $K$ (cf.~\cref{fig:afm-order-parameter}b). Unlike the paramagnetic insulator the spectral weight near the Fermi energy does not vanish (cf.~\cref{fig:para-spectral-fnc}). A broadening factor of $+i\eta = i0.025t_k$ was used because the self-energy $\Sigma(\omega)$ is real in the mean-field RISB approximation.
	}
\end{figure}

In \cref{fig:afm-dispersion} we show the spectrum of auxiliary fermions (\cref{eq:ham-qp}) for the antiferromagnetic insulator. The energy is lowered by opening a gap $\Delta$ at the high symmetry point $K$, which occurs because antiferromagnetic order breaks the inversion symmetry about the $t_g$ bond between triangles. In the paramagnetic solutions the Mott insulator occurred from a vanishing quasiparticle weight for the bands near the Fermi energy (\cref{sec:quasiparticle-bands}). In contrast, the quasiparticle weight in the antiferromagnetic insulator remains finite (\cref{fig:afm-spectral-fnc}). Instead, the broken symmetry opens a gap and there is a magnetic band insulator.

The antiferromagnetic insulator which occurs in the RISB theory of the decorated honeycomb lattice is crucially different to the description first proposed by Slater using Hartree-Fock theory. First, the SDW forms even though the lattice is not at half-filling. Second, the decorated honeycomb lattice is not bipartite and does not have perfect nesting. Hence, the insulator occurs at finite $U$ rather than for any $U>0$. Third, the spin-$1$ antiferromagnetic state is beyond a single-site Hartree-Fock description with ordering between sites. Instead, on the decorated honeycomb lattice spin-triplets form because of the effective Hund's interaction, with antiferromagnetic order between triangles. Fourth, usually a Slater insulator has a doubling of the unit-cell associated with the magnetic sublattice. However, because the SDW forms between triangles the magnetic sublattice has the same periodicity as the decorated honeycomb lattice.

\begin{figure}
	\centering
	\includegraphics[width=\columnwidth]{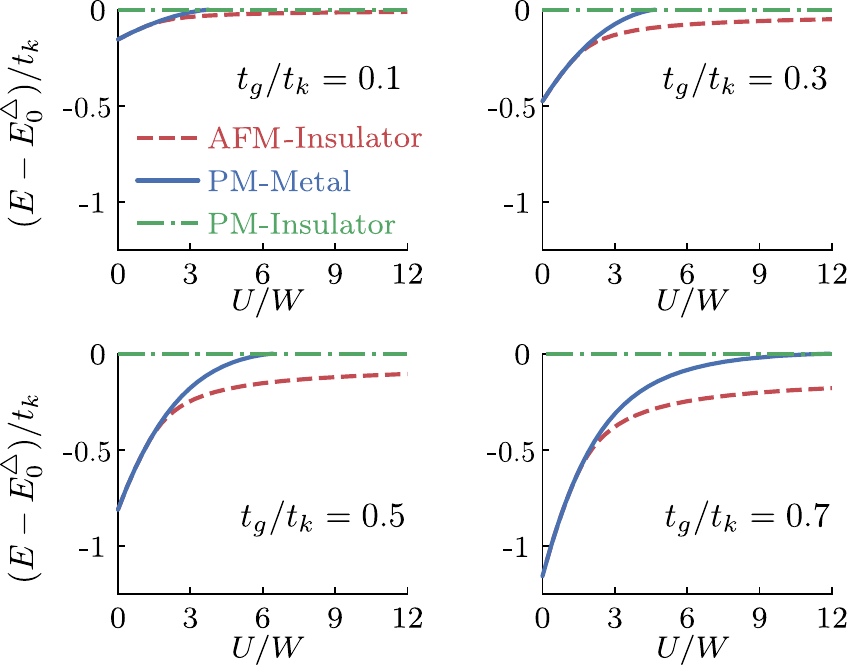}
	\caption{\label{fig:energy}
		Energy per triangle comparisons of the different solutions within RISB. The antiferromagnetic (AFM) insulator is the ground state for $U / W \gtrsim 1$ and occurs before the paramagnetic (PM) insulator. The energy of the PM insulator is equivalent to the ground state of an isolated triangle at two-thirds filling with $J_{\mathrm{GK}} = 0$, given by $E_0^{\triangle} = -2t_k + U$.
	}
\end{figure}

In \cref{fig:energy} we compare the energy of the antiferromagnetic insulator with the paramagnetic metal and the paramagnetic insulator. For $U / W \gtrsim 1$ the antiferromagnetic insulator has the lowest energy and is the ground state. The Slater insulator occurs at a lower critical interaction strength than the paramagnetic spin-$1$ Mott insulator.

We have only included results of the spin-1 antiferromagnetic insulator for $J_{\textrm{GK}} = 0$. This is in the regime where a spin-1 Mott insulator is found. However, we find that it is general that the spin-1 antiferromagnetic insulator occurs at a lower critical interaction strength than the spin-1 Mott insulator. The spin-1 antiferromagnetic insulator is never the lowest energy state in the regime where the spin-0 Mott insulator occurs.

However, we have only incorporated the antiferromagnetic order at the mean-field level, which ignores important quantum fluctuations that may suppress the formation of a SDW. It is also possible that there is a transition from the SDW to a N\'eel ordered state arising from the Mott insulator. Larger clusters of the lattice or incorporating fluctuations about the saddle-point \cite{Fabrizio2017,Lanata2017b,Wysokinski2017} may elucidate this issue.

\section{Low-energy effective theory for magnetism in the spin-$1$ Mott insulator} \label{sec:low-energy-theory}

The spin-$1$ Mott insulator on the decorated honeycomb lattice is adiabatically connected to the ground state of an isolated triangle: there are four electrons on each triangle that form a spin-triplet and inter-triangle charge and spin fluctuations vanish. However, the insulator described by RISB is an incomplete description of the true ground state of the spin-$1$ Mott state. Within the small cluster choice of our work, slave bosons ignore important higher order energy processes. Going beyond the slave boson approximation would include quantum fluctuations necessary to, e.g., capture the superexchange between triangles. 

Hence, we propose that the low-energy effective theory of the insulator is described by the antiferromagnetic spin-$1$ Heisenberg model on the honeycomb lattice \cite{Merino2016,Powell2017,Merino2017}, given by
\begin{equation} \label{eq:heis}
\hat{H}^{\mathrm{eff}} = J_1 \sum_{\langle ij \rangle} \vec{S}_i \cdot \vec{S}_j + J_2 \sum_{\langle \langle ij \rangle \rangle} \vec{S}_i \cdot \vec{S}_j,
\end{equation}
where we include a nearest-neighbor $J_1$ and next-nearest-neighbor $J_2$ coupling. From perturbation theory $J_1$ is given by \cite{Merino2016}
\begin{equation}
	J_1 = \frac{44 t_g^2}{81 U} + \frac{2 t_g^2 J_{\mathrm{GK}}}{81 t_k^2},
\end{equation}
and $J_2 / J_1 \approx 0.37 (t_g/U)^2$.
For $J_1 > 0$, $J_2 = 0$ the ground state of the spin-$1$ Heisenberg model on the honeycomb lattice is N\'eel ordered \cite{Gong2015,Li2016,Merino2018,Merino2018b}.

On a frustrated honeycomb lattice where $J_2/J_1 > 0$ the picture is less clear. A density matrix renormalization group (DMRG) study \cite{Gong2015} suggests a possible plaquette valence-bond crystal for $0.27 < J_2/J_1 < 0.32$ between the N\'eel ordered state and a spiral AFM, which is supported by a coupled cluster study \cite{Li2016} that finds a spin disordered state in the parameter range $0.25 < J_2/J_1 < 0.34$. In contrast, a study using Schwinger boson mean-field theory \cite{Merino2018b} observes no spin disordered state and instead finds a direct transition to the spiral AFM for $J_2/J_1 > 0.21$. However, Schwinger bosons may underestimate quantum effects \cite{Bauer2017,Merino2018b}. A quantum field theory suggests that the spin disordered region between the N\'eel state and spiral AFM is generic to two-dimensional lattices and that the quantum spin liquid shares similarities in its spin gap and spin fluctuations to the Haldane phase found in spin-$1$ chains \cite{Kharkov2018}. It is interesting that spin-$1$ formation also occurs in electronic models of quasi-one-dimensional lattices decorated with triangles, whose ground state is in the Haldane phase \cite{Janani2014a,Janani2014b,Nourse2016,Reja2019}. The nature of triplet formation of triangular structures in other decorated lattices and its relation to real materials remains an open question.

On the decorated honeycomb lattice the highest estimate of the frustration is $J_2 / J_1 \approx 0.03$ for $J_{\mathrm{GK}} = -J/3$ and a critical interaction strength $J / W \approx 1$ (\cref{fig:para-effective-J}). Hence, we predict that the ground state of the spin-$1$ Mott insulator on the decorated honeycomb lattice is N\'{e}el ordered.

\section{Conclusion} \label{sec:conclusion}

A spin-$1$ Mott insulator occurs on the Hubbard model on the decorated honeycomb lattice at two-thirds electron filling, where one would naively expect a correlated metal. The spin-$1$ Mott insulator can be driven by increasing electronic correlations or moving the lattice closer to the molecular limit (decreasing $t_g/t_k$). The resulting low-energy effective theory is the spin-$1$ Heisenberg model on the honeycomb lattice. While the decorated honeycomb lattice is not bipartite, the honeycomb lattice is, and the ground state of the spin-$1$ Mott insulator is N\'{e}el ordered.

An antiferromagnetic intra-triangle spin exchange drives the spin-$1$ Mott insulator to a spin-$0$ Mott insulator via a metallic phase. Both insulators are adiabatically connected to the limit of an isolated triangle. The spin-$1$ Mott insulator is a spin-triplet because of an effective Hund's coupling in the basis of trimer orbitals that favors high-spin on a triangle. The antiferromagnetic intra-triangle spin exchange suppresses the effective Hund's coupling and a spin-$0$ Mott insulator is favored instead.

The honeycomb aspect of the model is not crucial to realize the spin-$1$ and spin-$0$ Mott insulators. The insulators are governed by the local interactions in the molecular orbital picture of a triangle. We have previously demonstrated that decoration in lattices realizes ground states that are analogous to those found in multi-orbital models \cite{Nourse2021}. Because the spin-$1$ and spin-$0$ Mott insulators are adiabatically connected to the ground state of isolated triangles, the important feature of the lattice is that it is decorated with triangles where intra-triangle hopping dominates over inter-triangle hopping. The spin-$1$ and spin-$0$ Mott insulators may occur in other lattices decorated with triangles. However, the nature of the metal-insulator transition will likely differ because the Dirac metal occurs because of the honeycomb-like lattice.

For example, in quasi-one-dimensional electronic models spin-triplet formation on a triangle results in the Haldane phase, which is the ground state of the spin-$1$ Heisenberg chain \cite{Janani2014a,Janani2014b,Nourse2016,Reja2019}. We propose that spin-$1$ insulators is general to other lattices decorated with odd-sided polygons.

Doping the Haldane phase in quasi-one-dimensional models leads to spin-triplet superconducting fluctuations \cite{Reja2019}. Two questions arise from this observation. First, are the preformed triplets in the Haldane phase important for driving superconductivity, or does the unconventional superconductivity arise from the Mott insulator? Second, does spin-triplet superconductivity occur in higher dimensions of coupled triangles or is the low dimensionality important? In particular, it is an open question whether superconductivity occurs on the decorated honeycomb lattice in the vicinity of the spin-$1$ Mott insulator. It has already been shown that $f$-wave spin-singlet superconductivity can occur near half-filling on the decorated honeycomb lattice \cite{Merino2021}.

\begin{acknowledgments}
	We thank Jason Pillay and Elise Kenny for helpful conversations. This work was supported by the Australian Research Council through Grants No. DP160102425, DP160100060, and DP181006201.
\end{acknowledgments}

\appendix

\section{Hubbard $U$ in molecular orbital basis} \label{sec:app-local}

For any $N$-membered ring the single-particle transformation to the molecular orbital basis can be written as
\begin{equation}
\hat{c}_{i\alpha \sigma} \equiv \frac{1}{\sqrt{N}} \sum_m \hat{b}_{im\sigma}^{\dagger} e^{-i \phi (\alpha - 1) m},
\end{equation}
where $\phi = 2\pi / N$, and $m \in [0,N)$ is an index labeling the $N$ molecular orbitals. 

The on-site Coulomb repulsion in the molecular orbital basis is given by
\begin{equation}
\hat{H}_i^{U,N} \equiv U \sum_{\alpha} \hat{n}_{i\alpha \uparrow} \hat{n}_{i\alpha \downarrow} = \frac{U}{N} \sum_{ \mathclap{m n p} } \hat{b}_{im \uparrow}^{\dagger} \hat{b}_{in \uparrow}^{} \hat{b}_{ip \downarrow}^{\dagger} \hat{b}_{i(m + p - n) \downarrow}^{},
\end{equation}
where the identity for Fourier transforms $\frac{1}{N} \sum_{\alpha} e^{i\phi \alpha(n-m)} = \delta_{mn}$ was used, and hence $(m + n)$ is calculated modulo $N$. Expanding the terms on the right hand side gives
\begin{align} \label{eq:app-hubb-u-1}
& \hat{H}_i^{U,N} = \frac{U}{N} \sum_m \hat{n}_{im\uparrow} \hat{n} _{im\downarrow} \nonumber \\
& + \frac{U}{N} \sum_{ \mathclap{m\neq n} } \left( \hat{n}_{im\uparrow} \hat{n}_{in \downarrow} + \hat{b}_{im \downarrow}^{\dagger} \hat{b}_{in \uparrow}^{\dagger} \hat{b}_{im\uparrow}^{} \hat{b}_{in \downarrow}^{} \right) \nonumber \\
& + \frac{U}{N} \sum_{ \mathclap{m \neq n} } \hat{b}_{im\downarrow}^{\dagger} \hat{b}_{im\uparrow}^{\dagger} \hat{b}_{in\uparrow}^{} \hat{b}_{i(2m-n) \downarrow}^{} \nonumber \\
& + \frac{U}{N} \sum_m \sum_{n \neq n} \sum_{p \neq m,n}  \hat{b}_{i p \downarrow}^{\dagger} \hat{b}_{i m\uparrow}^{\dagger} \hat{b}_{i n \uparrow}^{} \hat{b}_{i (m + p -n) \downarrow}^{}.
\end{align}

The total spin operator can be written as
\begin{equation}
\vec{S}_i^2 = \sum_{m} \vec{S}_{im} \cdot \vec{S}_{im} + \sum_{ \mathclap{m \neq n} } \vec{S}_{im} \cdot \vec{S}_{in},
\end{equation}
with the total spin in an orbital given by
\begin{equation}
\vec{S}_{im}^2 \equiv \vec{S}_{im} \cdot \vec{S}_{im} = \frac{3}{4} \sum_{\sigma} \hat{n}_{im\sigma} - \frac{3}{2} \hat{n}_{im\uparrow} \hat{n}_{im \downarrow},
\end{equation}
and the spin exchange interaction between orbitals given by
\begin{align}
\vec{S}_{im} \cdot \vec{S}_{in} & = \frac{1}{4} \sum_{\sigma} \hat{n}_{im\sigma} \hat{n}_{in\sigma} - \frac{1}{2} \hat{n}_{im\uparrow} \hat{n}_{in \downarrow} \nonumber \\ 
& - \hat{b}_{im\downarrow}^{\dagger} \hat{b}_{in \uparrow}^{\dagger} \hat{b}_{im \uparrow}^{} \hat{b}_{in \downarrow}^{},
\end{align}
where we used that $\vec{S}_{im} = (S_{im}^x, S_{im}^y, S_{im}^z)$ is given by
\begin{equation}
	\vec{S}_{im} \equiv \frac{1}{2} \sum_{ \mathclap{\sigma \sigma' }} \hat{b}_{im \sigma}^{\dagger} \vec{\bm{\tau}}_{\sigma \sigma'} \hat{b}_{im \sigma'}^{},
\end{equation}
with $\vec{\bm{\tau}}$ a vector of Pauli matrices. The total number operator is given by
\begin{equation}
\hat{N}_i \equiv \sum_{ \mathclap{m\sigma} } \hat{n}_{im\sigma},
\end{equation}
and its square is given by 
\begin{align}
\hat{N}_i^2 & = \sum_m \hat{N}_i^2 + \sum_{ \mathclap{m\neq n} } \hat{N}_i^{2} \nonumber \\
& = \hat{N}_i + 2 \sum_m \hat{n}_{im\uparrow} \hat{n}_{im\downarrow} \nonumber \\
& + \sum_{ \mathclap{m\neq n, \sigma} } \hat{n}_{im\sigma} \hat{n}_{in\sigma} + 2 \sum_{ \mathclap{m\neq n} } \hat{n}_{im\uparrow} \hat{n}_{in\downarrow}.
\end{align}
Hence, the Hubbard $U$ interaction for an $N$-ring can be written in the molecular orbital basis as
\begin{align} \label{eq:N-cyclic-hubb}
& \hat{H}_i^{U,N} = - \frac{U}{N} \vec{S}_i^2 + \frac{U}{4N}  \hat{N}_i^2 + \frac{U}{2N}  \hat{N}_i - \frac{U}{N} \sum_m \hat{n}_{im\uparrow} \hat{n}_{im\downarrow} \nonumber \\
& + \frac{U}{N} \sum_{ \mathclap{m \neq n} } \hat{b}_{im\downarrow}^{\dagger} \hat{b}_{im\uparrow}^{\dagger} \hat{b}_{in\uparrow}^{} \hat{b}_{i(2m-n) \downarrow}^{} \nonumber \\
& + \frac{U}{N} \sum_m \sum_{n \neq n} \sum_{p \neq m,n}  \hat{b}_{i p \downarrow}^{\dagger} \hat{b}_{i m\uparrow}^{\dagger} \hat{b}_{i n \uparrow}^{} \hat{b}_{i (m + p -n) \downarrow}^{}.
\end{align}

For $N = 3$, the Hamiltonian for the Hubbard $U$ interaction in the trimer basis is given by \cref{eq-ham-U-trimer} in the main text.

\section{Spin exchange $J_{\mathrm{GK}}$ in molecular orbital basis} \label{sec:app-hunds}

The intra-triangle spin exchange interaction is given by
\begin{align}
\hat{H}_i^{F} & \equiv J_{\mathrm{GK}} \left( \sum_{\alpha \neq \beta} \vec{S}_{i \alpha} \cdot \vec{S}_{i \beta} - \frac{1}{4} \hat{n}_{i\alpha} \hat{n}_{i\beta} \right).
\end{align}
The first term can be written as
\begin{align}
\sum_{\alpha \neq \beta} \vec{S}_{i \alpha} \cdot \vec{S}_{i \beta} & = \vec{S}_i^2 - \sum_m \vec{S}_{im}^2 \nonumber \\
& = \vec{S}_i^2 - \frac{3}{4} \hat{N}_i + \frac{3}{2} \sum_{\alpha} \hat{n}_{i\alpha\uparrow} \hat{n}_{i\alpha \downarrow},
\end{align}
and the second term can be written as
\begin{align}
\frac{1}{4} \sum_{\alpha \neq \beta}  \hat{n}_{i\alpha} \hat{n}_{i\beta} = \frac{1}{4} \hat{N}_i^2 - \frac{1}{4} \hat{N}_i - \frac{1}{2} \sum_{\alpha} \hat{n}_{\alpha \uparrow} \hat{n}_{\alpha \downarrow}.
\end{align}
Hence, the spin exchange interaction within a triangle can be written as
\begin{align}
\hat{H}_i^{\mathrm{GK}} = J_{\mathrm{GK}} \vec{S}_i^2 - \frac{J_{\mathrm{GK}}}{4} \hat{N}_i^2 - \frac{J_{\mathrm{GK}}}{2} \hat{N}_i + 2 J_{\mathrm{GK}} \sum_{\alpha} \hat{n}_{i\alpha\uparrow} \hat{n}_{i\alpha\downarrow}.
\end{align}
Transforming the on-site Hubbard term using the results of \cref{sec:app-local} gives \cref{eq-ham-F-trimer} in the main text.

Note that combining the Hubbard $U$ and the spin exchange interaction gives
\begin{align}
& \hat{H}_i^U + \hat{H}_i^{\mathrm{GK}} = -\frac{1}{3}(U - J_{\mathrm{GK}}) \vec{S}_i^2 + \frac{1}{12}(U - J_{\mathrm{GK}}) \hat{N}_i^2 \nonumber \\
& + \frac{1}{6} (U - J_{\mathrm{GK}}) \hat{N}_i - \frac{1}{3}(U + 2 J_{\mathrm{GK}}) \sum_{m} \hat{n}_{im\uparrow} \hat{n}_{im\downarrow} \nonumber \\
& + \frac{1}{3}(U + 2 J_{\mathrm{GK}}) \sum_{m} \sum_{n \neq m} \sum_{p \neq m \neq n} \left( \hat{b}_{im \downarrow}^{\dagger} \hat{b}_{im \uparrow}^{\dagger} \hat{b}_{in \uparrow}^{} \hat{b}_{ip \downarrow}^{} + \mathrm{H.c.} \right).
\end{align}


%

\end{document}